# A Study of the Gas-Star Formation Relation over Cosmic Time[1]

R.Genzel[1,2], L.J.Tacconi[1], J.Gracia-Carpio[1], A.Sternberg[3], M.C.Cooper[4,5], K.Shapiro[6], A.Bolatto[7], N.Bouché[1,8], F.Bournaud[9], A.Burkert[10,11], F.Combes[12], J.Comerford[6], P.Cox[13], M.Davis[6], N.M. Förster Schreiber[1], S.Garcia-Burillo[14], D.Lutz[1], T.Naab[10], R.Neri[13], A.Omont[15], A.Shapley[16], & B.Weiner[4]

[1] Max-Planck-Institut für extraterrestrische Physik (MPE), Giessenbachstr.1, 85748 Garching, Germany

( linda@mpe.mpg.de, genzel@mpe.mpg.de )

[2] Dept. of Physics, Le Conte Hall, University of California, 94720 Berkeley, USA

[3] School of Physics and Astronomy, Tel Aviv University, Tel Aviv 69978, Israel

[4] Steward Observatory, 933 N. Cherry Ave., University of Arizona, Tucson AZ 85721-0065, USA

[5] Spitzer Fellow

[6] Dept. of Astronomy, Campbell Hall, University of California, Berkeley, 94720, USA

[7] Dept. of Astronomy, University of Maryland, College Park, MD 20742-2421, USA

[8] Dept. of Physics, University of California, Santa Barbara, Broida Hall, Santa Barbara CA 93106, USA

[9] Service d'Astrophysique, DAPNIA, CEA/Saclay, F-91191 Gif-sur-Yvette Cedex, France

[10] Universitätssternwarte der Ludwig-Maximiliansuniversität, Scheinerstr. 1, D-81679 München, Germany

[11] MPG-Fellow at MPE

[12] Observatoire de Paris, LERMA, CNRS, 61 Av. de l'Observatoire, F-75014 Paris, France

[13] IRAM, 300 Rue de la Piscine, 38406 St.Martin d'Heres, Grenoble, France

[14] Observatorio Astronómico Nacional-OAN, Apartado 1143, 28800 Alcalá de Henares- Madrid, Spain

[15] IAP, CNRS & Université Pierre & Marie Curie, 98 bis boulevard Arago, 75014 Paris, France

[16] Department of Physics & Astronomy, University of California, Los Angeles, CA 90095-1547, USA

[1] Based on observations with the Plateau de Bure millimetre interferometer, operated by the Institute for Radio Astronomy in the Millimetre Range (IRAM), which is funded by a partnership of INSU/CNRS (France), MPG (Germany) and IGN (Spain).

# Abstract

We use the first systematic data sets of CO molecular line emission in z~1-3 normal star forming galaxies (SFGs) for a comparison of the dependence of galaxy-averaged star formation rates on molecular gas masses at low and high redshifts, and in different galactic environments. Although the current high-z samples are still small and biased toward the luminous and massive tail of the actively star-forming 'main-sequence', a fairly clear picture is emerging. Independent of whether galaxy integrated quantities or surface densities are considered, low- and high-z SFG galaxy populations appear to follow similar molecular gas-star formation relations with slopes 1.1 to 1.2, over three orders of magnitude in gas mass or surface density. The gas-depletion time scale in these SFGs grows from 0.5 Gyrs at z~2 to 1.5 Gyrs at z~0. The average corresponds to a fairly low star formation efficiency of 2% per dynamical time. Because star formation depletion times are significantly smaller than the Hubble time at all redshifts sampled, star formation rates and gas fractions are set by the balance between gas accretion from the halo and stellar feedback.

In contrast, very luminous and ultra-luminous, gas rich major mergers at both low-z and high-z produce on average 4 to10 times more far-infrared luminosity per unit gas mass. We show that only some fraction of this difference can be explained by uncertainties in gas-mass or luminosity estimators; much of it must be intrinsic. A possible explanation is a top-heavy stellar mass function in the merging systems but the most likely interpretation is that the star formation relation is driven by global dynamical effects. For a given mass, the more compact merger systems produce stars more rapidly because their gas clouds are more compressed with shorter dynamical times, so that they churn more quickly through the available gas reservoir than the typical normal disk galaxies. When the dependence on galactic dynamical time scale is explicitly included, disk galaxies and mergers appear to follow similar gas to star-formation relations. The mergers may be forming stars at slightly higher efficiencies than the disks.

*Subject Headings: galaxies: evolution – galaxies: starbursts – galaxies: ISM – stars: formation – ISM: molecules*

# 1. Introduction

Stars form from neutral interstellar gas. In the Milky Way and nearby galaxies most and arguably all star formation occurs in massive ($10^4...10^{6.5}$ $M_\odot$), dense (n($H_2$)~$10^2...10^5$ cm$^{-3}$) and cold ($T_{gas}$~10-30 K), gravitationally bound 'giant molecular clouds' (GMCs: Solomon et al. 1987, Bolatto et al. 2008, McKee & Ostriker 2007). But how does star formation proceed on a global galactic scale? How does it depend on environmental parameters, such as the mass, density, temperature and kinematics of the interstellar gas, on the mass and properties of the galaxy on large scales, and on cosmic epoch? Is the mass function of newly formed stars (the initial mass function (IMF)) universal, or does it vary depending on environment? In large part because the microphysics of stellar formation itself is still not fully understood, these questions, central to quantitative models of galaxy evolution, cannot yet be answered by ab-initio theory (McKee & Ostriker 2007).

Ever since the pioneering work of Schmidt (1959), empirical scaling relations have been fairly successful in describing galactic scale star formation in z=0 disk galaxies. The simplest hypothesis is to relate the star formation rate (*SFR*) integrated over a galaxy, SFR, and its total neutral gas mass $M_{gas}$ through

$$SFR = M_{gas}/\tau_d \qquad (1),$$

where $\tau_d$ is the gas consumption/depletion time scale (Leroy et al. 2008, Bauermeister, Blitz & Ma 2010). The virtue of equation (1) is that it is easily accessible to global measurements of the tracers of star formation and gas (i.e. stellar luminosity, CO 1-0 line luminosity and HI mass) in a large number of galaxies (Young & Scoville 1991, Solomon & Sage 1988, Gao & Solomon 2004). Its disadvantage is that it probably is too simplistic.

A popular physical approach has been to assume that the star formation rate volume density, $\dot{\rho}_*$, scales with gas volume density $\rho_{gas}$ and local free-fall time $\tau_{ff}$ as (Schmidt 1959, Kennicutt 1998a (henceforth K98a), Krumholz & McKee 2005, Leroy et al. 2008)

$$\dot{\rho}_* = \varepsilon_{ff} \frac{\rho_{gas}}{\tau_{ff}} \propto \rho_{gas}^{1.5} \text{ since } \tau_{ff} \propto \rho_{gas}^{-1/2} \qquad (2),$$

where the (dimensionless) star-formation efficiency $\varepsilon_{ff} = \tau_{ff}/\tau_d$ is the fraction of the available gas mass converted to stars in a free-fall time (e.g. Krumholz & McKee 2005). Since volume densities cannot be easily determined on galactic scales, but surface densities can, the corresponding surface density relation is (K98a)

$$<\dot{\Sigma}_*> \propto \varepsilon_{ff}\left(<\Sigma_{gas}>\right)^N \quad (3),$$

where for constant vertical scale height, equation (3) follows from equation (2) with N=1.5 (K98a, Krumholz & McKee 2005, Leroy et al. 2008). Brackets indicate spatial averages across the galaxy. Alternatively one can set (Elmegreen 1997, Silk 1997, K98a)

$$<\dot{\Sigma}_*> = \varepsilon_\tau \frac{<\Sigma_{gas}>}{\tau_{dyn}(R_{gal})} \quad (4),$$

where $\varepsilon_\tau$ is the star formation efficiency per galaxy dynamical time. In a marginally stable galactic disk, the Toomre (1964) Q-parameter, $Q = \kappa\sigma/(\pi G \Sigma_{gas})$, is near unity. Here G is the gravitational constant, $\kappa$ and $\sigma$ are the epicyclic frequency ($\kappa \approx \xi \tau_{dyn}^{-1}$ ($\xi$~1.4...2)) and the local 1-d sound speed or velocity dispersion. For Q=1

$$\tau_{dyn} = \frac{\xi\sigma}{\pi G \Sigma_{gas}} \quad (5),$$

so that with $\Sigma_{gas} = \left(\rho_{gas-midplane}\sigma\tau_{dyn}\right)$ it can be easily seen that the free fall time scale of a gas cloud in the mid-plane, $\tau_{d,ff} = \left(\sqrt{\pi G \rho_{gas-midplane}}\right)^{-1}$ is approximately equal to the galaxy's global dynamical time scale, and $\varepsilon_{ff} \approx \varepsilon_\tau$. Leroy et al. (2008) have discussed other approaches that are variants of equations (1)-(4).

The basic empirical finding in the Milky Way and in local disk galaxies is that the global gas depletion times in z~0 disk galaxies are about 2 Gyrs (Leroy et al. 2008, Bigiel et al. 2008), and the global star formation efficiency is 1-3% per free fall, or dynamical time (K98a, Leroy et al. 2008). In his seminal 1998 paper Kennicutt finds N=1.4±0.15 from galaxy integrated measurements of 97 z=0 galaxies. This sample consists of 61 normal spirals and 36 infrared-bright starburst galaxies, and includes 5 luminous (LIRG: $\geq 10^{11}$ L$_\odot$) or ultra-luminous (ULIRG: $\geq 10^{12}$ L$_\odot$) merging galaxies. Bouché et al. (2007) compared the K98a sample to the first spatially resolved data at z~1-3. They concluded that low- and high-z galaxies follow a similar star formation relation but find a somewhat steeper slope than K98a (N~1.7), mainly because of a more appropriate CO to gas mass conversion factor for z~0 ULIRGs and z~1-3 submillimetre galaxies (SMGs) (e.g. Downes & Solomon 1998).

Krumholz & McKee (2005) have argued that the small star formation efficiencies are plausibly accounted for by the supersonic turbulent motions in GMCs (Mach numbers ~10-30) creating very broad (log-normal) gas density distributions. At any given time only a few percent of the gas is dense enough to collapse by self-gravity before it gets dispersed again by the turbulent motions. An analytic derivation (Krumholz, McKee &

Tumlinson 2009) including internal stellar feedback and external pressure leads to a relation similar to equation (2) but with a power law index N that varies from ~0.8 at low surface densities to ~1.3 at high densities (Krumholz et al. 2009). In this model the transition occurs at $\Sigma_{gas} \sim 10^2$ $M_\odot pc^{-2}$ below which the GMC pressure is dominated by internal star formation feedback, while external pressure dominates at higher surface densities. Recent spatially resolved studies of the star formation in nearby disk and irregular galaxies support the view that no single exponent N can account for the entire relationship (Bigiel et al. 2008, Leroy et al. 2008). At very low surface densities most of the gas is in atomic form and N is near 2(±0.5) if the relation between total gas (sum of atomic and molecular) and star formation is considered. Above a critical surface density of ~10 $M_\odot pc^{-2}$ most of the gas becomes molecular (Blitz & Rosolowsky 2006) and the relationship attains a slope near unity (Bigiel et al. 2008, Leroy et al. 2008). A comparison of these newer spatially resolved data to the original integrated starburst sample of K98a suggests that above ~$10^2$ $M_\odot pc^{-2}$ the relation may steepen again, with a slope >1 (Bigiel et al. 2008).

A specific issue is the inferred gas depletion time scale for infrared luminous merging galaxies. Evidence has accumulated since the mid-1980s that the ratio of infrared luminosity to CO line luminosity is much larger in interacting galaxies (by a factor of ~4-6, Young et al. 1986) and in z~0 ULIRG mergers (by a factor >10, Sanders, Scoville & Soifer 1991, Sanders & Mirabel 1996, Gao & Solomon 2004) than in normal spiral disks. In the interacting/merging systems gas surface densities are also much larger than in normal disk galaxies (Braine & Combes 1993, Downes & Solomon 1998, Tacconi et al. 2008). The interpretation of these findings is uncertain, however. Sanders et al. (1991) proposed that most z~0 ULIRGs are powered by AGN, which dominate the far-infrared (FIR) luminosity and thus lead to an overestimate of $L_{FIR}/L_{CO}$ as far as star formation is concerned. Gao & Solomon (2004) find that in contrast to CO 1-0, there exists a linear relationship between the dense gas tracer HCN 1-0 and $L_{FIR}$ in z~0 ULIRGs, which favours a shorter gas depletion time scale in luminous mergers as the cause of the FIR excess. Gao & Solomon (2004) propose that the underlying reason for this boost in $L_{IR}/M_{mol-gas}$ ratio is an increased fraction of very dense molecular gas in mergers.

Over the last decade CO line emission studies have become possible also for high-redshift galaxies. Investigations concentrated initially on very luminous quasars (Omont et al. 1996, Walter et al. 2003, Greve et al. 2005) and submillimetre galaxies (Frayer et al. 1998, 1999, Genzel et al. 2003, Greve et al. 2005, Tacconi et al. 2006, 2008, Chapman et al. 2008). Because of significant advances in the sensitivity of the IRAM Plateau de Bure millimetre interferometer, integrated and spatially resolved CO measurements have very recently become feasible also in 'normal' z~1-2.5 SFGs (i.e. not predominantly major mergers and not extreme starbursts: Daddi et al. 2008, 2010, Dannerbauer et al. 2009, Tacconi et al. 2010). These high-z SFGs are situated on the 'main-sequence' in the star formation rate-stellar mass plane, and appear to have high-duty cycles (30-60%) indicative of near-continuous star formation and replenishment of new gas (Noeske et al. 2007, Daddi et al. 2007, Bouché et al. 2010). Less than 50% of high-z SFGs are major mergers (Shapiro et al. 2008, Förster Schreiber et al. 2009, Tacconi et al. 2010, Daddi et al. 2010). Although the high-z samples are still small and biased toward the massive tail of the main sequence (stellar masses 3-20x$10^{10}$ M$_\odot$), it is obviously of great interest to compare the star formation relation in these systems relative to their z=0 counterparts, as well as to the low- and high-z luminous mergers. This comparison is the subject of the present paper. All physical units used below are based on a concordance, flat ΛCDM cosmology with $H_0$=70 km s$^{-1}$ Mpc$^{-1}$, $\Omega_m$=0.28, $\Omega_\Lambda$=0.72. Stellar masses and star formation rates are based on a Chabrier (2003) initial stellar mass function (IMF).

## 2. Sample Selection and Properties

In this paper we analyse data sets from a number of studies. Most importantly we discuss the recent data on z~1-2.5 normal, massive star forming galaxies reported in Tacconi et al. (2010), Daddi et al. (2010) and Tacconi et al. (in prep.), all obtained with the IRAM Plateau de Bure millimetre Interferometer. The Tacconi et al. (2010) study (with the addition of 3 galaxies in Tacconi et al. in prep.) reports on two mass- and star formation-matched samples of <z>=1.2 and <z>=2.3 SFGs, with currently 10 galaxies at <z>=1.2 and 11 galaxies at <z>=2.3, and with the same criteria of stellar mass (>4x$10^{10}$ M$_\odot$) and star formation (>70 M$_\odot$yr$^{-1}$) selection. Daddi et al. (2010) have observed a sample of 6 SFGs at <z>=1.5, with comparable selection criteria as in Tacconi et al. For a description of the observations and the data analysis we refer to the papers above. We compare these recent observations with measurements of z~1-3.5 submillimeter galaxies, and several data sets on z~0 normal, star bursting and merging galaxies from the literature. For all galaxies we adopt the K98b relation between star formation rates and infrared (8-1000µm luminosity), $SFR$ (M$_\odot$/yr)=$10^{-10}$ $L_{IR}$ (L$_\odot$).

In the following we discuss the properties of these different data sets.

### 2.1 <z>=1.2 sample

The z~1 SFGs discussed in this paper (10 galaxies) are drawn from AEGIS (Davis et al. 2007, Noeske et al. 2007). The All-Wavelength Extended Groth Strip International Survey (AEGIS) provides deep imaging in all major wave bands from X-ray to radio (including Advanced Camera for Surveys (ACS) HST images), as well as optical spectroscopy (DEEP2/Keck) over a large area of sky (0.5 deg$^2$), with the aim of studying the panchromatic properties of galaxies over the last half of the Hubble time. The region studied is the Extended Groth Strip (EGS: RA=14$^h$17$^m$, Dec= 52$^0$30'). The AEGIS data provide the properties of a complete set of galaxies from 0.2≤z≤1.2 for the stellar mass range >$10^{10}$ M$_\odot$. Extinction corrected star formation rates are derived from a combination of Spitzer MIPS 24µm fluxes, GALEX UV fluxes and Hα/[OII] fluxes (Noeske et al. 2007, Cooper et al. in prep.). Direct observations far-infrared spectral energy distributions (SEDs) of z~1 SFGs with luminosities comparable to our EGS-galaxies have now become possible with PACS on the Herschel[2] space telescope. The initial observations yield similar far-infrared luminosities as estimated from data at other wavelengths (Elbaz et al. 2010). From the AEGIS data set we selected SFGs at z~1.1 to1.3 without obvious major merger morphologies, with stellar masses ≥$4\times10^{10}$ $M_\odot$ and with star-formation rates ≥70 $M_\odot$ yr$^{-1}$. The EGS galaxies thus represent the most luminous tail of the star forming <z>=1.2 galaxy population. Their star formation rates place them on the 'main sequence' in the $M_*$-SFR plane (Noeske et al. 2007) and at the high mass tail of the overall population (left panel in Figure 1). Table 1 lists the inferred star formation rates and stellar masses, as well as half-light radii. The latter are obtained from Sersic fitting to the HST I/K-band images, with the exception of EGS13035123, EGS1207881 and EGS13003805, where the quoted radii are averages of these Sersic parameters and fits to the spatially resolved CO 3-2 distributions. In these three galaxies the half-light radii obtained from CO and stellar light are in reasonable agreement, similar to the finding for z=0 disk galaxies (Young & Scoville 1991, Leroy et al. 2008).

---

[2] See http://herschel.esac.esa.int/SDP_wkshops/presentations/IR/6_Lutz_PEP_SDP2009.pdf

Most of the <z>=1.2 SFGs in our sample are rotating disks. The CO kinematics in the 3 spatially resolved galaxies discussed above clearly indicate rotational motions in large disks with big local gas/star forming clumps (Tacconi et al. 2010, and in prep). In three additional galaxies velocity gradients are detected, which are indicative of rotational motions as well. The morphologies of the ACS images also are fully consistent with a clumpy disk interpretation. The values of the maximum disk circular velocity $v_d$ quoted in Table 1 were computed from the velocity difference in the two line profile emission peaks, or from the velocity width and corrected for inclinations obtained from the minor to major axis ratios in the I-band images. For details we refer to Tacconi et al. (2010).

## 2.2 <z>=2.3 sample

The z~2.3 SFGs (11 galaxies) were selected from the near-infrared long-slit Hα sample of Erb et al. (2006), which in turn was drawn from the larger survey of Steidel et al. (2004) and Reddy et al (2005), culled according to the so-called 'BX' criteria based on UV-colour (UGR) and R-magnitude (hereafter, simply BX sample). Our sub-sample was chosen from these surveys to cover the same stellar mass and star formation range as the z~1.2 AEGIS sample. The right panel of Figure 1 shows that within the measurement uncertainties and given the intrinsic scatter of the z~2 'main sequence', the 'BX' sample provides a fair census of the high mass end of the entire UV-/optically selected SFG population in this red-shift and mass range (Reddy et al. 2005, Erb et al. 2006, Förster Schreiber et al. 2009). As for the z~1 sample, many of the massive z~2 SFGs are turbulent, clumpy rotating disks (Cresci et al. 2009). Less than 50% of the massive z~2 SFGs galaxies studied with high-resolution Hα integral field spectroscopy exhibit kinematical properties that would be expected for major mergers (Shapiro et al. 2008, Förster Schreiber et al. 2009).

The star formation rates listed in Table 1 were derived from extinction corrected Hα-luminosities (Erb et al. 2006, Förster Schreiber et al. 2009). For this purpose the conversions in Kennicutt (1998b (K98b)) were applied but the 'Salpeter IMF' star formation rates were divided by 1.7 to convert to a Chabrier IMF. Stellar masses are from the spectral energy distribution fits in (Erb et al. 2006, Förster Schreiber et al. 2009), and $v_d$ and $R_{1/2}$ values are from the data in the same references with the methods discussed in (Förster Schreiber et al. 2009). The typical uncertainties of the derived quantities are dominated in most sources by systematic errors in stellar masses, star formation rates and $H_2$ masses, all of which are at least 50%.

## 2.3 <z>=1.5 sample

In addition, we included in our analysis 6 <z>=1.5 SFGs from Daddi et al. (2008, 2010). Four of these 6 sources have spatially resolved PdBI observations. All four of these resolved galaxies show strong velocity gradients with double-peaked profiles, indicative of ordered rotational motions in large disk galaxies, similar to our BX/EGS samples (Daddi et al. 2010). The ACS images of these galaxies exhibit large clumps, again similar to what is observed at z~2. These BzK galaxies have comparable stellar masses (~$5 \times 10^{10}$ $M_\odot$) and star formation rates (~150 $M_\odot yr^{-1}$). The star formation rates reported by Daddi et al. (2010) are based on several star formation indicators. They average estimates from the extinction corrected UV luminosities, from the radio luminosities, and from extrapolations of the observed 24μm Spitzer mid-infrared (rest-frame ~10μm) luminosities to far-infrared luminosities with the library of z=0 Chary & Elbaz (2001) spectral energy distributions. We estimate that the luminosities and implied star formation rates are uncertain by at least 50% (for a given IMF). Stellar masses were estimated from optical/UV SED fitting in the same way as at z~1 and 2, with similar uncertainties. Half-light radii are averages of the values obtained from CO and ACS.

## 2.4 <z>=1-3.5 SMG sample

Finally we also included 20 z=1-3.5 SMGs from the PdBI SMG survey presented in Greve et al. (2005), Tacconi et al. (2006, 2008), Engel et al. (2010) and Smail et al. (2010, in prep.). For 10 of these SMGs we have spatially resolved data. All but two of these SMGs are drawn from the luminous tail of the SMG population with $S_{850\mu m}$>4 mJy, corresponding to $L_{FIR}$>$4.4 \times 10^{12} L_\odot$ and SFR>440 $M_\odot yr^{-1}$ when applying the Pope et al. (2006) relation between 850μm flux density and FIR luminosity, and the K98b relation between far-infrared luminosity and star formation rate (corrected to the Chabrier IMF). Direct far-infrared observations of SMGs in GOODS-N with Herschel/PACS (see footnote 2; Magnelli et al. 2010) confirm the Pope calibration. The stellar masses of the four SMGs with good quality SED fitting are >$10^{11} M_\odot$ (Tacconi et al. 2008). The kinematics and structure of the SMGs are quite different from the z~1-2.5 SFGs. Only a few of the ~10 SMGs with high resolution PdBI CO imaging exhibit evidence for ordered motions and several consist of binary systems in close interaction. Their size distribution and 'messy' kinematic properties are best explained by a scenario where most of these luminous SMGs are powerful starbursts triggered by major gas rich binary mergers in different phases of the merging process, similar to z~0 ULIRGs (Engel et al. 2010).

Table 1 summarises all relevant observed and derived properties of the spatially resolved high-z SFGs and SMGs. To compute surface densities we divided half of the star formation rate or gas mass by the area subtended by the half-light (or effective) radius ($\pi R^2$).

## 2.5 <z>~0 samples

We compare these high-redshift star forming galaxy samples to several z~0 star forming galaxy samples from the literature. We have taken normal disk galaxies from the original K98a paper and from Leroy et al. (2008, 2009; henceforth 'z~0 normal') and Kuno et al. (2007).. We compiled a list of non-merging starburst galaxies from K98a, Gao & Solomon (2004), Kuno et al. (2007) and Gracia-Carpio (2009; henceforth z~0 'SFGs'). A sample of z~0 ULIRGs and interacting galaxies was constructed from the samples of K98a, Gracia-Carpio et al. (2008) and Gracia-Carpio (2009). For galaxies with multiple observations we either selected the measurement with the highest signal-to-noise ratio, or we selected observations sampling different spatial scales. We did not use observations of galaxy nuclei only, in an attempt to screen against the impact of AGN. We also only included galaxies with molecular surface densities above 3 $M_\odot pc^{-2}$, where the molecular to atomic gas fractions become comparable or greater than 1 (Bigiel et al. 2008). Our final z~0 data base has ~150 entries. All data were transformed to the same cosmology and conversion factors as given in the Introduction.

## 2.6 Conversion factors from CO luminosity to molecular gas mass

Observations in Milky Way GMCs have established that the integrated line flux of $^{12}$CO millimetre rotational lines can be used to infer cold (molecular) gas masses, despite the fact that the CO molecule only makes up a small fraction of the entire gas mass and that the lower rotational lines (1-0, 2-1, 3-2) are almost always very optically thick (Dickman, Snell & Schloerb 1986, Solomon et al. 1987). This is because the CO emission in the Milky Way and nearby normal galaxies on average comes from moderately dense (volume averaged densities $<n(H_2)> \sim 200$ cm$^{-3}$), self-gravitating GMCs. In this regime the ratio of molecular hydrogen column density to line integrated CO intensity, or of molecular gas mass (including a 36% mass correction for helium) to CO luminosity $L'_{CO}$ ($L'_{CO} = \int_{source} \int_{line} T_R(v) \, dv \, dA$ [K km/s pc$^2$]) can be expressed as

$$N(H_2)/I(CO) = X = c_1 \left( \frac{\sqrt{<n(H_2)>}}{T_R} \right), \quad [\text{cm}^{-2}/(\text{K km s}^{-1})] \quad (6)$$

and

$$M_{gas}/L'_{CO} = 1.36 \cdot \alpha = c_2 \left( \frac{\sqrt{<n(H_2)>}}{T_R} \right), \quad [M_\odot / (\text{K km s}^{-1} \text{pc}^2)] \quad (7).$$

Here $T_R$ is the equivalent Rayleigh-Jeans brightness temperature of the (optically thick) CO line and $c_1$ and $c_2$ are appropriate numerical constants. In the 2.6mm CO (1-0) transition the typical gas temperature of GMCs ranges from 10 to 25 K. Several independent empirical techniques based on GeV γ-rays, optical extinction measurements, isotopomeric line ratios and excitation analysis have all shown that this 'virial' technique is appropriate and remarkably robust throughout the Milky Way (Dickman et al. 1986, Solomon et al. 1987, Strong & Mattox 1996, Dame, Hartmann & Thaddeus 2001). The best empirical Galactic conversion factor is $X_G \sim 2 \times 10^{20}$ [cm$^{-2}$ (K km s$^{-1}$)$^{-1}$] and $\alpha_G = 3.2$ [M$_\odot$ (K km/s pc$^2$)$^{-1}$]. The virial approach can be shown to also apply to an ensemble of virialized clouds, instead of a single one, as long as again the factor $n(H_2)^{1/2}/T$ is approximately constant throughout the system and the CO line is optically thick (Dickman et al. 1986).

For the z~1-2 SFGs a Galactic conversion factor ($\alpha \sim 3.2$ M$_\odot$/(K km s$^{-1}$ pc$^2$)) is appropriate since the CO emission in these systems, as in z~0 disk galaxies, probably arises in virialized giant molecular cloud systems (GMCs) and mean gas densities of $<n(H_2)> \sim 10^{2...3}$ cm$^{-3}$ (Tacconi et al. 2010). Based on far-infrared SEDs from PACS/SPIRE Herschel (see footnote 2; Elbaz et al. 2010) and an observation of several CO rotational lines in one galaxy (Dannerbauer et al. 2009) these SFGs may have gas/dust temperatures of 20-35 K. For the large column densities (and interstellar pressures) in the z>1 SFGs ($\Sigma_{gas} >> 10$ M$_\odot$ pc$^{-2}$) most of the cold interstellar gas is probably in molecular form and the contribution of atomic hydrogen can be neglected (Blitz & Rosolowsky 2006).

The assumption of an ensemble of individual gas clouds in virial equilibrium with their own gravity breaks down, however, in galactic nuclei and mergers. In these cases gas velocities are dominated by disturbed large scale motions in the gravitational potential of gas, stars and dark matter, and the gas is often in a smoother, disk-like configuration, rather than in virialized individual clouds. In this limit the relationships above still hold but with a modified proportionality factor (Solomon et al. 1997, Scoville, Yun & Bryant 1997, Downes & Solomon 1998, Sakamoto et al. 1999). For conditions appropriate for z~0 ULIRGs ($n(H_2) \sim 10^3$-$10^4$ cm$^{-3}$ and $T_R \sim 20$-60 K) with gas fractions of ~20%, the inferred empirical conversion factor ranges from $X_{merger} = 5 \times 10^{19}$ (or $\alpha_{merger} \sim 0.8$: Solomon et al. 1997, Downes & Solomon 1998) to $10^{20}$ ($\alpha_{merger} \sim 1.6$, Scoville, Yun & Bryant 1997). A Galactic conversion factor can be excluded for these systems as in that case the gas fractions would significantly exceed unity. Furthermore, the assumption of a Galactic conversion factor for these luminous mergers would imply that the two progenitor galaxies also had anomalously high molecular gas masses relative to normal spiral galaxies of the same stellar mass. An analysis of gas, stellar and dynamical masses suggests that the local ULIRG/merger conversion factor is also appropriate for very luminous z>1 SMGs (Tacconi et al. 2008). If a Galactic conversion is applied the ratio of gas masses to dynamical masses becomes much greater than unity.

In the z≥1 SFGs and SMGs rotationally excited (J>1) transitions are observed and corrections for the ratio of the intrinsic Rayleigh-Jeans brightness temperatures in the 1-0 line to that in the rotationally excited line may be necessary. Everything else being equal and all levels being thermalized, the difference between Planck ($T_{Planck}$) and Rayleigh-Jeans brightness temperatures ($T_R = \dfrac{h\nu/k}{\exp(h\nu/kT_{Planck}) - 1}$) leads to corrections between 1.15 (1.23, 1.33) and 1.33 (1.55, 1.82) for temperatures between 40 and 20 K for the J=2-1 (J=3-2, J=4-3) transitions. These are lower limits since the upper states may not be thermalized. CO line ratios in z~0 disk galaxies (Mauersberger et al. 1999) and in the z~1.5 SFG BzK21000 (Dannerbauer et al. 2009) suggest the following correction factors for SFGs: $R_{1J}=L'_{CO\ 1-0}/L'_{CO\ J-(J-1)}$~1.2 and 2 for J=2 and 3, respectively. For the SMGs in Table 1 we use $R_{1J}$=1.1, 1.3 and 1.6 for J=2,3 and 4 as motivated by observed line ratios in SMGs (Weiss et al. 2007) and z~0 ULIRGs (Iono et al. 2009).

# 3. Luminosity-Luminosity Correlation

The first and most straightforward approach, independent of the choice(s) of conversion factor(s) from CO line luminosity to $M_{mol-gas}$, is to investigate the relation between the observed (z=0) or inferred (z≥1) far-infrared luminosities and the observed (z=0) or inferred (z≥1) CO 1-0 luminosities. The results are shown in Figure 2. Compared to other plots discussed below, the total (statistical plus systematic) uncertainties are smallest in this plot. We estimate typical 1σ uncertainties of ±0.13 dex (30%) and ±0.17 dex (40%) for $L_{CO}$ and $L_{FIR}$, respectively. In the case of the z≥1 SFGs and SMGs direct observations of the far-infrared luminosities with the Herschel satellite confirm the applied calibrations to get $L_{FIR}$ from the original star formation tracers (UV, Hα, mid-infrared, submillimeter, radio: see footnote 2; Magnelli et al. 2010, Elbaz et al. 2010), or give corrections. Nordon et al. (in press.) find that at z~1.5-2.5 24μm based star formation rates typically overestimate the directly measured far-infrared luminosities by a factor of four.

The data and plots shown in Figure 2 are in excellent agreement with the results of all past studies on the subject (e.g. Sanders et al. 1991, Sanders & Mirabel 1996, Gao & Solomon 2004, Greve et al. 2005, Gracia-Carpio et al. 2008). They confirm that there is a close to linear (slope 1.15±0.12) relation between $L_{CO}$ and $L_{FIR}$ for SFGs (a constant gas depletion time scale) and that the luminous mergers lie on average a factor of about 4 above the relation for the 'normal' star forming galaxies. All uncertainties for slopes and offsets given henceforth and in the Figures are 3σ formal fit errors, which is a reasonable estimate of the total error, including systematic effects, calibration etc.

Our analysis adds two important new aspects. The first is that the gas depletion time for SFGs estimated from the right panel of Figure 2 depends only weakly on redshift. The redshift averaged ratio $L_{FIR}/L_{CO\ 1-0}$ is 27±5.6 $L_\odot$/(K km/s pc$^2$) for SFGs. The correlation has a fairly substantial dispersion of 0.31 dex (see K98a). There are also several outliers. The CO luminosities in EGS12012083 and BX389 are surprisingly weak despite an estimated luminosity of >$10^{12}L_\odot$. To get from far-infrared to total infrared luminosity we use $L_{IR}$(8-1000μm)/ $L_{FIR}$(50-300μm)~1.3 (Gracia-Carpio et al. 2008). Taking the K98b conversion for SFR/$L_{IR}$ and α=$α_G$=3.2 for SFGs (section 2.6) this results in an effective gas consumption time scale for SFGs of 1.2 (±0.3) Gyr averaged from z=0 to 2. If the sample is split by redshift, then for the z=0 SFGs the gas depletion time is 1.5 Gyrs and for the z≥1 SFGs it is 0.5 Gyrs. For comparison Leroy et al. (2008) find a value of 1.9±0.9 Gyrs for their z=0 disk sample. The difference between our z~0 time scale and that of Leroy et al. is the correction factor $L_{IR}/L_{FIR}$ mentioned above. The gas depletion time scale for the z≥1 SFGs is comparable to their 'stellar ages', as estimated either from the ratio of stellar mass and star formation rate, or from synthesis modelling of their UV- to near-infrared restframe SEDs (Förster Schreiber et al. 2009).

The second important point relates to the interpretation of the offset between SFGs and luminous mergers. In past studies there were no SFGs above ~$10^{10}$ K km/s pc$^2$, such that the difference between mergers (z~0 ULIRGs and z≥1 SMGs) and normal z=0 disk galaxies was interpreted by most observers as a luminosity effect, with a change of slope occurring above ~$10^{10}$ K km/s pc$^2$ (or $10^{11.5}$ $L_\odot$), in the merger regime (Bouché et al. 2007). With the addition of the z≥1 SFGs this interpretation may not be tenable anymore. Instead it now appears that the locations of mergers and SFGs on average are well separated at the same CO and IR luminosities. Normal (disk) galaxies with star formation occurring over long time scales and with a high duty cycle constitute one extreme. Luminous major mergers, predominantly after the first encounter and later stages of the merging process (Veilleux, Kim & Sanders 2002, Engel et al. 2010) and in an extreme (maximum starbursts: Tacconi et al. 2006) and brief starburst phase (duty cycle ~10%: Tacconi et al. 2008) constitute the other. Luminous mergers have an average gas depletion time of 0.2±0.11 Gyrs, 2.5 to 7.5 shorter than the SFGs (right panel of Figure 2). Note that the somewhat lower luminosity interacting z~0 starbursts in our sample (filled cyan squares in Figure 2), which are not in a late stage of merging, are located closer to the isolated SFGs, and not near the extreme mergers.

SFGs and luminous mergers each adhere to a near-linear relation but the respective relations are offset by 0.6 dex. Such an offset has also been seen by other authors when comparing local ULIRGs (Gracia-Carpio et al. 2008) and z~2 SMGs (Bothwell et al. 2010) with local SFGs. For a fixed slope the probable uncertainty in this offset of the mean is ±0.15 and the offset is significant. However, the scatter of both merger and SFG distributions is large (~0.31 dex) at both low- and high-z and there is significant overlap between the two populations, as well as outliers. We come back to the interpretation of this offset in section 6.

# 4. Surface Density Correlations

## 4.1 SFGs

We now consider the distribution of the SFGs in the classical 'Kennicutt-Schmidt' surface density plane. Our best estimate for the typical 1σ total uncertainties (statistical plus systematic) are ±0.23 dex (52%) and ±0.27 dex (62%) for $\Sigma_{mol\,gas}$ and $\Sigma_{star\,form}$, respectively. Figure 3 summarises the distribution for the low- and high-z SFGs. As in the case of the pure luminosity plot in Figure 2 the correlation of all data has a standard deviation relative to the best fit of 0.32 dex. Again this scatter is somewhat larger than the typical total measurement uncertainties shown as a cross in the lower right of the left panel of Figure 3. The slope of the relation is 1.17 (±0.09), where the errors for slope-fits quoted here and below are formal 3σ fit errors. Note that in our fitting in Figure 3 (and also Figure 2) we did not attempt to assign individual errors (unlike K98a), since in our opinion essentially all uncertainties are systematic in nature and apply to all data equally. This slope is in very good agreement with the spatially resolved relation for nearby spirals in Bigiel et al. (2008, green/orange/red-shaded region in the left panel of Figure 3). The new data do not indicate a significant steepening of the slope at surface densities of $>10^2$ $M_\odot pc^{-2}$, neither at z~0 nor at z≥1. Within the limited statistics of the currently available data, we do not find a break in the slope near $10^2$ $M_\odot$ $pc^{-2}$, as proposed by Krumholz et al. (2009). The slope of 1.33 found by Krumholz et al. (2009) in the high density limit is marginally larger. A steeper slope in this regime (1.28 to 1.4) was suggested earlier by the K98a starburst sample, but that analysis included some mergers (see below) and the combined scatter of both data sets suggest a 1σ uncertainty of ~0.15, which makes the difference in slope of 0.1-0.23 only marginally significant.

Low- and high-z SFGs overlap completely, again with the obvious exception of EGS12012083 and BX389. The data in Figure 3 suggest that the KS-relation in normal star forming galaxies does not vary with redshift, in agreement with the conclusions of Bouché et al. (2007) and Daddi et al. (2010).

In the right panel of Figure 3 we analyse the data with the 'Elmegreen-Silk'-relation (see also K98a), which relates star formation rate surface density to the ratio of gas surface density and global galaxy dynamical timescale. There is a reasonably good correlation as well with a slope of slightly less than unity (0.84±0.09). The scatter in this relation (0.44 dex) is larger than in the surface density relation, which may in part be attributable to the larger total uncertainties in $\Sigma_{mol\,gas}/\tau_{dyn}$, which we estimate to be ±0.32 dex (74%). Here and elsewhere we computed the dynamical time scale from the ratio of the radius to the circular velocity $v_c$. For the z>1 SFGs and SMGs we took R=$R_{1/2}$ and applied a pressure correction to the inclination corrected rotation velocity $v_{rot}$, $v_c=(v_{rot}^2 + 2\sigma^2)^{1/2}$, where σ is the local 1d-velocity dispersion in the galaxy. This relation is applicable to rotation-dominated, as well as pressure dominated galaxies. The slope we find is close to that of K98a, who find a slope between 0.9 and 1. High-z SFGs have somewhat higher $\Sigma_{star\,formation}$ than low-z galaxies (by 0.71±0.21 dex) but the difference is probably only marginally significant. A fit with unity slope yields a star formation efficiency per dynamical time of 0.019 (±0.008). This is in excellent agreement with 0.01, the value found by K98a when corrected to a Chabrier IMF.

## 4.2 KS-relation for Luminous Mergers

Figure 4 summarises our analysis of the luminous mergers at both low- and high-z. The left panel shows the case of applying the best single common conversion factor determined from the observations ($\alpha_{merger} \sim 1$, section 2.6), such that mergers and SFGs now have conversion factors that differ by a factor of 3.2. The slope of the merger relation (1.1±0.2) is consistent with that of the SFGs (1.17). Again low- and high-z mergers lie plausibly on the same relation. Independent of whether the merger slope is fit or forced to be the same as that of the SFGs, the difference in star formation rate at a given gas surface density between the two branches is ~1.0 (±0.2) dex (see also Bothwell et al. 2010).

As we have argued in section 2.6, a Galactic conversion factor for all luminous low- and high-z mergers is almost certainly excluded (Downes & Solomon 1998, Sakamoto et al. 1999, Tacconi et al. 2008). If one, nevertheless, adopts the same conversion factor as for 'normal' galaxies ($\alpha_{merger} = \alpha_{SFG} = 3.2$), the mergers move by ~0.5 dex to higher gas mass surface densities (right panel of Figure 4), closer to but not fully overlapping with the relation for SFGs. A fit to the entire data set then yields a steeper slope of 1.27 (±0.075) for equal weighting.

## 4.3 Elmegreen-Silk relation for mergers and SFGs

In Figure 5 we added the low- and high-z luminous mergers in the $\Sigma_{mol\,gas}/\tau_{dyn}$-$\Sigma_{star\,form}$-plane. As discussed in section 4.1, the scatter in this plane is generally somewhat larger (±0.31 dex for mergers, ±0.45 dex for SFGs and ±0.55 dex for all data combined). The mergers are still 0.5 to 0.7 dex above the SFGs, suggesting a correspondingly greater star formation efficiency per dynamical time scale $\varepsilon_\tau$ (equation 4). However, the offset is less than in the gas surface density-star formation surface density plane. Given the larger intrinsic scatter, all data may thus be drawn from a single underlying distribution, albeit with a larger scatter than in the relation for either SFGs or mergers by themselves. The scatter may be driven by uncertainties, especially in the estimation of dynamical time, or by (a) additional (hidden) parameter(s). The best fit for all data combined has a slope consistent with unity (0.98, ±0.09), similar to K98a, and a zero point offset of -1.76 (±0.18). This offset corresponds to a star formation efficiency of 1.7% per dynamical time scale. The physical interpretation of this common underlying relation then would be that the star formation relation is in part driven by the large scale dynamical time in a system. Spiral density waves and/or galaxy interactions drive density waves and bars, which in turn trigger cloud formation, enhanced cloud-cloud collisions and gas compression and global radial streaming (see Leroy et al. 2008). Mergers are more luminous because of a combination of their smaller sizes and dynamical time scales (Table 1, Downes & Solomon 1998) and an additional larger efficiency of star formation per dynamical time scale. This is indeed the interpretation proposed by Bouché et al. (2007). Similar conclusions follow when considering galaxy integrated quantities (SFR as a function of $M_{mol-gas}/\tau_{dyn}$, see section 6.3).

## 4.4 Comparison to K98a and Bouché et al. (2007)

We finish here with a brief discussion of the differences in slopes found in different studies. As discussed several times above, the slope determinations in the KS-surface density relation range between N~1.0 (Bigiel et al. 2008, Leroy et al. 2008, section 4.1) and N=1.7 (Bouché et al. 2007), with K98a and Kennicutt et al. (2007) in the middle (N=1.3 to 1.4). How can one understand the fairly large differences in slope, given that each of these papers quotes (1σ) uncertainties of ±0.15 or less?

A first important factor is the definition of gas surface densities. Throughout this paper we have discussed the 'molecular' gas-star formation relations and have restricted ourselves to $\Sigma_{mol\,gas} > 3\ M_\odot pc^{-2}$ where molecular gas should begin to dominate the ISM (Blitz & Rosolowsky 2006). The value of N=1.0 for Bigiel et al. (2008) is also for the molecular relation at higher surface densities. K98a and Bouché et al. (2007) consider the total cold gas column densities, including a contribution from HI. This contribution is plausibly negligible for z~0 starbursts, ULIRGs and z≥ SFGs and SMGs but not for normal galaxies with $\Sigma_{mol\,gas} < 10\ M_\odot pc^{-2}$. It is easy to understand that leaving out the HI contribution and only considering the molecular gas tends to decrease surface densities at the lower end, and thus flattens the distribution. If the K98a data are re-analyzed without the HI contributions but leaving all other assumptions the same as in K98a, we find N=1.33 instead of N=1.4. If the HI-columns are included for the normal galaxies in our SFG sample, the slope in Figure 3 would change from 1.17 to 1.28.

The second important factor is the choice of conversion factor. A constant (Galactic) conversion factor for all galaxies, as assumed in K98a, yields a flatter distribution than does the bimodal value we favour. The molecular KS-relation for the K98a data reanalyzed in this way yields N=1.42 (compared to N=1.33 above).

Finally, the choice and weighting of data also significantly affects the resulting slopes. By adding a number of z≥1 SMGs at high surface densities, combined with the fairly extreme choice of a bimodal conversion factor ($\alpha_{merger}$=0.25$\alpha_{SFG}$) and an inclusion of the HI contribution at the low surface density end, Bouché et al. (2007) arrived at the most extreme result in the literature of N=1.7 (see also Bothwell et al. 2010). K98a and Kennicutt et al. (2007) show that by different weightings of data and by using data with different spatial resolutions, the slope shifts by about ±0.1. Bigiel et al. (2008) average the spatially resolved molecular KS-relation for 18 z~0 normal galaxies and find N=1.0±0.2, while Kennicutt et al. (2007) find N=1.37±0.03 for the spatially resolved relation in M51 only. The slope of N=1.17 derived for the SFGs in Figure 3 is in part driven by a number of new normal and starburst galaxies we have included from the recent literature. These galaxies were not in the K98a sample and tend to increase the number of galaxies at $\Sigma_{mol\,gas} \sim 10^{2.8...3.8} M_\odot pc^{-2}$ and push the fit to a somewhat shallower slope.

We conclude that the slope of the KS-surface density relation is sensitive to a number of subtle systematic effects, each of which can change the derived fit-value by ±0.1. The total systematic uncertainty of slope determinations probably is ±0.2 to ±0.25. When comparing different results in the literature care needs to be taken to understand which assumptions were made to be able to assess the reality of the differences in quoted values.

## 4.5 Summary of the observations

In summary of the sections 3 and 4, we find

- molecular gas and star formation rates (in terms of total quantities or surface densities) in normal star forming galaxies are correlated with a slope of 1.1 to 1.2 across the entire range observed ($10^{0.5}$ to $10^4$ $M_\odot pc^{-2}$). The scatter is fairly large (0.32 dex) but arguably significantly affected by systematic uncertainties;
- low- and high-redshift samples follow similar relations. This means that in normal star forming galaxy populations with a semi-continuous star formation history the star formation efficiency in SFGs ever since ~ 3 Gyrs after the Big Bang has been low (1.7% per dynamical time), or the gas depletion time scale long, ~60 times the dynamical time. The molecular gas depletion time throughout this period has increased slowly from 0.5 at z~2 to 1.5 Gyrs at z~0;
- luminous starburst-mergers at both low and high redshift appear to produce 4 to 10 times more far-infrared luminosity per gas mass than normal SFGs. Apart from a zero-point offset, the relations for merger driven systems are characterized by essentially the same slope as the "normal" SFGs;
- the difference between mergers and SFGs is minimized in the $SFR - M_{mol-gas}/\tau_{dyn}$-plane or the $\Sigma_{mol\,gas}/\tau_{dyn}$-$\Sigma_{star\,form}$-plane, albeit at the cost of a larger scatter than in the luminosity and surface density-planes.
- The star formation efficiencies per dynamical time may be somewhat larger in mergers than in SFGs.

# 5. Discussion of Uncertainties

## 5.1 AGN Contamination

For all galaxies in Figure 2 the total (infrared) luminosity is a measure of the star formation rate, unless there is a substantial contamination by an AGN. Most of the SFGs are probably not strongly affected. This is by selection, since with one exception (NGC 1068) the z~0 SFG sample excludes bright AGN. In NGC1068, which is used here, the AGN contributes about half of the infrared luminosity (Telesco et al. 1984). In several well-known z~0 ULIRGs, such as Mrk231 and Mrk273, the far-infrared emission of the central AGN likely dominates the luminosity, in others it probably contributes a fraction (≤50%) of the luminosity (Sander & Mirabel 1996, Genzel et al. 1998, Veilleux et al.2009). On average, the intrinsic z~0 ULIRG star formation luminosities could be overestimated in Figures 2 and 4 by a factor of 1.5 to 2. For most of the z≥1 SMGs, AGNs are present but the AGN contribution to $L_{FIR}$ is probably relatively small (≤ a few tens of percent: Lutz et al. 2005, Alexander et al. 2005, Valiante et al. 2007, Menendez-Delmestre et al. 2007, Pope et al. 2008). To minimize the possible contamination by AGNs we have taken the far-infrared luminosity and multiplied by 1.3 to estimate the total (8-1000μm) infrared luminosity associated with star formation, $L_{IR}$. This approach avoids using the mid-infrared luminosity that is more prone to AGN contamination (Netzer et al. 2007). This correction factor is the average of $L_{IR}/L_{FIR}$~1.3 in nearby normal star forming galaxies (Gracia-Carpio et al. 2008). The z≥1 SMG luminosities given in Figures 2 and 4 thus could also result in overestimates of their star formation luminosities by a modest factor (1-1.5).

## 5.2 From star formation tracers to SFRs

Different observers have used different star formation tracers, with each method having its own strength and weakness (see e.g. K98b, Kennicutt et al. 2009). The Hα recombination line luminosity is an excellent measure of the instantaneous formation rate of the most massive (O) stars (Figure 5), but requires a good knowledge of the extinction correction. In practice, the Hα indicator is very well suited for low extinction star forming galaxies, including most normal z~0 disk galaxies and probably all but the dustiest z≥1 SFGs. It tends to fail in the most extreme, dusty SFGs and z~0 ULIRGs/ z≥1 SMGs because of the very large extinctions (e.g. Goldader et al. 1995, Genzel et al. 1998). The far-UV luminosity and the far-infrared luminosity (a proxy of bolometric luminosity in dusty SFGs) measure the integral of star formation over a fairly long period of time and thus require a good knowledge of the star formation history (Figure 5). The far-UV luminosity is even more sensitive to extinction than Hα, and for this reason it fails in the case of very dusty star forming galaxies (Goldader et al. 2002). Since the extinction of high-z SFGs appears to be lower than comparably luminous low-z systems (Daddi et al. 2007, Reddy & Steidel 2009), the applicability of the UV-estimator (like that of Hα) appears well justified for most z≥1 SFGs.

With the advent of very sensitive measurements from the Spitzer Space Telescope for studying high-z SFGs, it has become increasingly popular to use the observed 24μm luminosity (corresponding to 6-12μm in the rest frame), in conjunction with a library of z=0 SFG SEDs (e.g. Chary & Elbaz 2001), for an extrapolation to the far-infrared luminosity. Direct observations of FIR emission of z≥1 SFGs and SMGs with PACS on the Herschel Observatory indicate that the 24μm-extrapolation method is robust at z≤1.5, but overestimates FIR luminosities in z>1.5, L >$10^{12}$ $L_\odot$ SFGs and SMGs by factors of 2-4 (footnote 2; Elbaz et al. 2010, Nordon et al. 2010). For 3 of the 6 z~1.5 BzK galaxies studied by Daddi et al. (2010), the PACS-FIR luminosities are comparable to those derived from the 24μm data, and on average no corrections to the SFRs are necessary (Nordon et al. 2010)

Several authors are using combinations of the various tracers to overcome their respective disadvantages (Daddi et al. 2007, Noeske et al. 2007, Calzetti et al. 2007, Kennicutt et al. 2007, 2009). The luminosities in our <z>~1.2 sample are based on such combinations, and thus should be fairly secure. For the <z>=2.3 SFGs we use the observed Hα luminosities from Erb et al. (2006) and Förster Schreiber et al. (2009) and apply the Calzetti et al. (2000) prescription, with the proposed extra nebular attenuation relative to stellar light by 2.3, to estimate the Hα-extinction from the E(B-V) values in Erb et al. (2006). A comparison of these Hα-luminosities/star formation rates, extinction corrected far-UV luminosities/star formation rates and direct PACS-Herschel far-infrared luminosities/star formation rates yields encouraging agreement (Förster Schreiber et al. 2009, Nordon et al. 2010). This agreement instils some trust in the applicability of the empirical Calzetti et al. (2000) extinction curve, perhaps also including the extra factor of 2.3 for the ratio of nebular to continuum extinction, for high redshift SFGs (see discussion in Förster Schreiber et al. 2009, but see Reddy et al. 2010).

Once one has obtained a best estimate intrinsic far-infrared/UV-/Hα-luminosity and has ascertained that it is powered by star formation, the next step is to go from luminosity to star formation rate. This step involves a star formation history and an IMF. Figure 56 shows the dependence of the various tracers on star formation history. It is clear that the various standard conversions from these tracer luminosities to SFR require that star formation does not vary rapidly and preferably is constant (see K98b for a discussion). Hα, far-IR and far-UV luminosities all tend to underestimate the peak SFR for a starburst past its maximum, in some cases by large factors, unless the star formation history can be explicitly included in the modelling. Even for a constant star formation history, the infrared to SFR conversion used in Figure 4 ($L_{IR}=10^{10}$ SFR) applies only for ages $>>10^7$ years. Hα as an estimator of instantaneous star formation does better but is obviously strongly dependent on the exact shape of the IMF in the O-star mass range (>20 $M_\odot$).

## 5.3 From $L_{CO\ 1-0}$ to $M_{mol-gas}$

We have discussed in section 2.6 the basic assumptions and reasoning for being able to convert integrated CO 1-0 line luminosities to total molecular gas masses with a simple proportionality factor. This conversion is fundamentally fraught with several systematic uncertainties. Yet studies with a range of methods have come up with similar conversion factors in the Milky Way and nearby near-solar metallicity disk galaxies, to within a factor of ≤2 (see the Appendix in Tacconi et al. 2008 for a more detailed discussion). Because of photo-dissociation of the molecular gas, the situation in low metallicity (<50% solar) dwarf systems is more uncertain, although even there a Galactic conversion appears to apply in the dense star forming clumps themselves (Bolatto et al. 2008). Our database does not include these lower metallicity systems, and most of the more massive z≥1 SFGs probably have near-solar metallicities (Erb et al. 2006b).

We have emphasized in section 2.6 the important assumption that the molecular line emission originate in self-gravitating clouds where gas dominates the mass balance. This is because the basic mass information is contained in the observed line width (virial estimate) and not in the intensity (equation 6, Dickman et al. 1986, Solomon et al. 1987). This assumption is reasonably well justified in isolated disk galaxies close to thermal and dynamical equilibrium. It probably also holds in many star bursting systems, as long as the integrity of the GMCs is maintained, and the factor $\sqrt{<n(H_2)>}/<T_{gas}>$ is approximately the same as in the Milky Way (0.5-1.5). It arguably is also applicable to the z~1-2 SFGs since in the few cases studied so far, their properties are close enough to the Milky Way (presence of giant clouds, gas densities and temperatures) that a similar conversion factor is plausible (Tacconi et al. 2010, Daddi et al. 2010, Dannerbauer et al. 2009).

The situation is quite different in the merging galaxies. While there still may be a reasonably well defined conversion factor 3 to 4 times smaller than the Galactic value, its value is probably uncertain by a factor of at least 2. In some instances, such as the gaseous bridge between the two nuclei of NGC 6240, the gas motions may be completely out of equilibrium and the concept of a virial mass estimator based on line width may not be applicable (c.f. Tacconi et al. 1999). In general, z~0 ULIRGs and z≥1 SMGs are highly disturbed systems with large velocity dispersions, and the application of dynamical tracers, such as virial estimators or rotation curves, may yield uncertain (factor ≥2) or even questionable results (Mihos 1999).

We have chosen the simplest approach of assigning a single constant conversion factor each for mergers and SFGs. The reality undoubtedly is more complex. The conversion factor may be a function of physical parameters, such as surface density of gas and star formation, metallicity etc.(Obreschkow & Rawlings 2009). There may be some support for such a smooth conversion function as a function of gas surface in Figure 10 of Tacconi et al. (2008) but the trend is not well enough defined to apply it here. The effect of introducing such a function would be for luminous (high surface density) SFGs to have lower conversion factors, more akin to mergers. The gap between mergers and SFGs in Figures 2 and 4 would then be less pronounced and the overall distribution would be more continuous, and not as bimodal as Figure 4 suggests. There would then be a steepening of the overall gas-star formation rate relation at $>10^3$ $M_\odot pc^{-2}$, with a slope ~1.4, consistent with the interpretation of Bouché et al. (2007). This does not change the basic conclusion, however, that luminous mergers on average have a larger $L_{FIR}/M_{gas}$ ratio than SFGs.

### 5.4 Spatially resolved vs. integrated measurements

If star formation in a given galaxy is a strong function of position because of the presence of a radial gradient in the Toomre Q-parameter, spiral arms, bar-induced resonances, nuclear concentrations and large star forming complexes in the disk, galaxy integrated studies may yield a different gas-star formation relation than spatially resolved studies. Because galaxy integrated measurements smear out such structures one would expect galaxy integrated relations to be shallower than spatially resolved data. This is in fact observed in M51, where Kennicutt et al. (2007) find N=1.56±0.04 in 520 pc data, N=1.37±0.03 in 1850 pc resolution data, and N=1.4±0.15 for the global average data (when compared to other galaxies). Undoubtedly such effects will be also present in the z≥1 SFGs, which have a very clumpy and irregular star formation and gas distributions (Genzel et al. 2008, Tacconi et al. 2010). Future spatially resolved studies of the gas-star formation relations in the z≥1 SFGs will shed light on this issue.

# 6. Possible Interpretations

## 6.1 cosmic evolution of the gas-star formation relation in SFGs

The near-unity slope of the gas-star formation relations and the slow variation of the gas depletion time scale as a function of cosmic time in normal SFGs are at first glance surprising. Bigiel et al. (2008) have interpreted the near-linear relation between $\dot{\Sigma}_*$ and $\Sigma_{mol-gas}$ in low-z disks in terms of the galaxy-scale filling factor of the basic building blocks, namely star forming GMCs. Since GMC properties in normal z~0 galaxies are probably quite comparable (Bolatto et al. 2008), a similar linear relation in different galaxies is plausible. The interstellar medium in z≥1 SFGs is quite different from that in the local Universe, however. The turbulent velocities and Mach numbers in z≥1 SFGs are far greater than at z~0 ($\sigma_{gas}$~20-80 km/s instead of 5-10 km/s, Förster Schreiber et al. 2009, Cresci et al. 2009, Tacconi et al. 2010). Gas column densities and pressures are also much larger (Elmegreen 2009). As a result the Toomre-mass, that is the largest gravitational unstable mode of gas structure formation, is about $10^2$ greater, and feedback probably less destructive (e.g. Elmegreen 2009, Dekel, Sari & Ceverino 2009, Genzel et al. 2008). If the global star formation relation were strongly sensitive to these internal properties of the star forming complexes, one might have expected to see a difference in slope or an offset for the high-z SFGs. The fact that the gas-star formation relation apparently is not strongly altered by these rather different internal properties thus suggests to us that the global gas depletion time scale (or efficiency per free fall/dynamical time scale) is set or strongly influenced at large scales, as assumed in the Elmegreen-Silk relation, rather than or in addition to small-scale processes in star forming clumps (e.g. Struck et al. 2005). Alternatively the star formation relation could, in principle, be sensitive to the internal properties but their effects may be compensated by self-regulation.

Another important feature is the slow change in gas depletion time from 0.5 to 1.5 Gyrs over 10 Gyrs, which is smaller than the Hubble time for all redshifts observed here. At the same time the SFGs sampled by the current observations have large star formation rates (e.g. Noeske et al. 2007, Daddi et al. 2007, Franx et al. 2008) and high gas fractions (30-60%, Tacconi et al. 2010, Daddi et al. 2010) throughout the z~1-2.5 redshift range. Taken together these facts require that the gas reservoir is replenished semi-continuously over cosmic time (Tacconi et al. 2010, Bouché et al. 2010, Bauermeister et al. 2010). Bauermeister et al. (2010) have shown that the observed change in depletion time scale with cosmic time can be understood as the consequence of the variations in gas accretion and star formation. This conclusion is consistent with other observational (e.g. Erb 2008, Förster Schreiber et al. 2006, 2009, Genzel et al. 2008) and theoretical findings (Dekel & Birnboim 2006, Kereš et al. 2005, Ocvirk, Pichon & Teyssier 2008, Dekel et al. 2009), which all suggest a semi-continuous supply of fresh gas in 'cold flows' from the intergalactic cosmic web, below a critical halo mass of ~$10^{12}$ $M_\odot$. Since the gas depletion time scale in SFGs is not a strong function of cosmic epoch, star formation rates and gas fractions as a function of red-shift are likely set by the balance between accretion from the halo and feedback (Bouché et al. 2010).

## 6.2 what accounts for the excess in $L_{FIR}/L_{CO}$ in mergers?

What is the interpretation of the excess in $L_{FIR}/L_{CO}$ and $\dot{\Sigma}_* - \Sigma_{mol-gas}$ in z~0 ULIRGs and z≥1 SMGs? Sanders et al. (1991) proposed that AGNs strongly dominate the energy output of ULIRG mergers. In the meantime ISO and Spitzer spectroscopy, as well as Chandra X-ray observations have made a compelling case that most z≥1 SMGs and the majority of z~0 ULIRGs are powered predominantly (but by no means exclusively) by star formation (Genzel et al. 1998, Alexander et al. 2005, Veilleux et al. 2009). The AGN contribution may on average reduce the excess of 4-10 by a factor of 2 or less but unlikely will remove it.

As long as only submillimetre data were available for z≥1 SMGs, one possible explanation for the strong emission of SMGs was to postulate that they are much colder than expected, thereby lowering the required far-infrared luminosities substantially (Kaviani, Haehnelt & Kauffmann 2003, Efstathiou & Rowan-Robinson 2003). The results emerging from PACS on Herschel probably exclude that possibility. Magnelli et al. (2010) find that the SEDs of luminous z≥1 SMGs are very similar to the standard interpretations (Chapman et al. 2005, Pope et al. 2006), with 30 to 40 K dust temperatures. z≥1 SMGs are indeed very luminous systems.

The uncertainties in going from star formation tracers to star formation rates (given an IMF) are substantial but have now been mitigated by combining different star formation tracers and having available direct far-infrared luminosities from Herschel for the luminous high-z populations (SFGs as well as SMGs, see footnote 2; Magnelli et al.

2010, Nordon et al. 2010, Elbaz et al. 2010). A definite concern is the step from $L_{FIR}$ to SFR, which requires the assumption of a star formation history. Most SFR calibrations assume steady star formation histories over $>>10^7$ years, which are a poor description for the bursting merger systems. However, we have shown that in this case the correction would go in the direction of increasing the intrinsic excess in the ratio of SFR/$M_{gas}$ in the merging systems relative to the SFG populations.

The conversion factor from CO luminosity to gas mass is quite uncertain in mergers. However, the excess in the merger data already is strongly present in the $L_{FIR}/L_{CO}$ relation, which does not make any assumptions about a conversion factor. The discussion in section 4.2 has shown that any reasonable application of an improved conversion factor applicable for mergers makes the excess stronger, not weaker. A Galactic conversion factor (which would remove the discrepancy at least in the surface density plane) is in all likelihood excluded in most mergers because of mass balance. We thus conclude that improved estimates of the conversion factor, perhaps in form of a conversion function dependent on a number of physical parameters, may perhaps lessen and smear out the difference between mergers and SFGs, but will not completely remove it.

This leaves in our opinion two options. If the IMF in mergers would be significantly more top-heavy than in normal disks, the $L_{FIR}/M_{mol-gas}$ ratio would increase proportionally. This option has already been proposed for other reasons by Baugh et al. (2005), Davé (2008) and van Dokkum (2008). The trouble with that explanation is that there is no convincing local Universe evidence (yet ?), other than in the immediate vicinity of the black hole in the Galactic Centre (Bartko et al. 2010), that significantly top-heavy IMFs are found under physical conditions similar to major mergers (Bastian, Covey & Meyer, 2010). Studies of super-massive star clusters in star burst galaxies and nearby mergers are perhaps the most promising hunting ground, but the observational support for a top-heavy in these cases is marginal at best (de Grijs & Parmentier 2007, Bastian et al. 2010).

The most likely explanation thus is that luminous mergers are forming stars faster (shorter depletion time scale) and/or more efficiently than normal disks at any cosmic epoch. Plausible qualitative explanations might appeal to large scale shocks for compressing the gas and increasing the fraction of dense gas that can then collapse and form stars. However, major mergers are also more turbulent than SFGs (Mach numbers in ULIRGs are ~200, compared to ~25 in Milky Way GMCs). Turbulent dissipation time scales thus are correspondingly smaller as well. The star formation efficiency per free fall/dynamical time may even decrease with increasing Mach number (Krumholz & McKee 2005).

## 6.3 the global dynamical time scale is an important parameter

The discussions in section 4.3 and in 6.1 point to the large scale, dynamical time scale as a critical additional element in the global star formation relations at low- and high-redshifts and in different mass, gas fraction and dynamical time regimes. Figure 7 shows what happens if the dynamical time scale is added as an explicit parameter in the 3d-space spanned by $\tau_{dyn}$-SFR-$M_{mol\text{-}gas}$. In the $M_{mol\text{-}gas}$-SFR (similar to Figure 2) and $\tau_{dyn}$-SFR projections the four different galaxy types (z~0 and z≥1 SFGs, and luminous mergers) clearly separate, as discussed throughout this paper. However, there exists a projection of the three-dimensional distribution, in which mergers and SFGs at both low- and high-redshift follow a well-defined relation of modest scatter (±0.47 dex) given by

$$\log(SFR\ (M_\odot yr^{-1})) = -0.78(\pm 0.23)\log(\tau_{dyn}(yr)) + 1.37(\pm 0.16)\log(M_{mol-gas}(M_\odot)) - 6.9(\pm 1.9) \quad (8).$$

The uncertainties given in equation (8) again are 3σ fit errors, Note that the -0.8-slope dependence of SFR on $\tau_{dyn}$ can be easily seen in the sequence between SMGs and z≥1 SFGs, which plausibly sample galaxies of similar gas, stellar and dynamical masses (Tacconi et al. 2008, 2010). The detailed values of the slopes determined from the data naturally depend strongly on the weighting of the different types of galaxies in the extremes of the distribution. The values given in equation (8) are for equal weights to z~0 SFGs, z≥1 SFGs, z~0 mergers (ULIRGs) and z≥1 mergers (SMGs). Equation (8) describes a 'fundamental plane' of star formation. For the relation above the slope in the $M_{mol\text{-}gas}$-SFR projection is similar to K98a and the theoretical work of Krumholz & McKee (2005) and Krumholz et al. (2009). The slope in the $\tau_{dyn}$-SFR projection is only marginally different from a simple SFR~$M_{mol\text{-}gas}/\tau_{dyn}$ relation, which was discussed in section 3. Future work is needed to determine whether this subtle tilt of the relation is real and what it might be caused by.

The basic interpretation is that the more compact z≥1 SMGs and z~0 ULIRG mergers form stars more rapidly because their gas clouds are more compressed with shorter dynamical times, so that they churn more quickly through the available gas reservoir than the typical normal disk galaxies. When the dependence on galactic dynamical time scale is explicitly included, disk galaxies and mergers appear to follow similar gas to star-formation relations. The mergers may be forming stars at slightly higher efficiencies than the disks. This interpretation, also favoured by Bouché et al. (2007), is consistent with our remarks earlier in this section about the remarkable uniformity of z~0 and z≥1 SFGs, despite obvious physical differences in their star forming clouds. If the global star formation efficiency per free fall/dynamical time scale is driven by large scale dynamical effects, such as spiral arms, global disk fragmentation and galaxy scale compression, then a single gas-star formation relation may broadly explain both the mergers and SFGs in our sample.

# 7. Conclusions

We have analysed in this paper galaxy-integrated measurements of molecular gas and star formation rates in galaxies spanning a wide range of properties. We have included z~0 normal star forming galaxies, very luminous and gas rich, but otherwise normal star forming disks at z~1-2.5, z~0 ULIRG mergers, and z≥1 submillimetre galaxies, many of which are also gas rich, dissipative mergers.

To extract from the basic data (CO line luminosities and far-infrared/submillimetre or UV luminosities) the underlying physical parameters, namely molecular gas masses and star formation rates, we have used the best available consensus conversion and calibration factors. While these calibrations are undoubtedly approximate and have substantial uncertainties, we find that our main findings are quite robust and do not change qualitatively over the plausible range of these calibrations.

Our first main conclusion is that the gas-star formation relation does not depend much on redshift. Star formation in both z~0 and z~1-2.5 normal star forming disks is quite inefficient. The gas depletion time (~1.5 Gyr at z~0 and ~0.5 Gyr at z~1-2.5) is ~50 times greater than average local free-fall, or galaxy-scale dynamical times. However, the gas depletion time is also much shorter than the Hubble time at all redshifts. A semi-continuous replenishment of gas through smooth accretion or minor mergers is thus required.

The ratio of star formation rate to available molecular gas mass in ULIRGs and SMGs is four to ten times greater than in SFGs at all redshifts. Since z~1-2.5 SFGs are as luminous as z~0 ULIRGs and almost as luminous as z≥1 SMGs, this L/M-excess cannot be a pure luminosity effect. If the global galactic dynamical time scale is introduced as an explicit third parameter, all galaxies appear to lie in a plane, with a scatter that is only somewhat greater than the systematic measurement uncertainties. The difference between SFGs, ULIRGs and SMGs disappears almost completely in the edge-on projection of this plane, which may be somewhat tilted with respect to $45^0$ (which would correspond to SFR$\propto M_{mol-gas}/\tau_{dyn}$). We propose that major gas rich mergers form stars more efficiently and rapidly because of the combination of the more rapid driving of large scale gas compression and the higher fractions of dense gas. Our second major conclusion thus is that global star formation in normal disk galaxies as well as the most extreme galactic environments can be largely captured in a 'universal' gas-star formation relation explicitly including the dynamical time.

# References

Alexander D. M., Bauer F. E., Chapman, S. C., Smail I., Blain A. W., Brandt W. N., Ivison R. J., 2005, ApJ, 632, 736
Bartko H., Martins F., Trippe S., et al., 2010, ApJ, 708, 834
Bastian N., Covey K. R., Meyer M.R., 2010, ARA&A, in press (arXiv:1001.2965)
Bauermeister A., Blitz L., Ma C.-P. 2010, ApJ, submitted (astro-ph 0909.3840)
Baugh C. M., Lacey C. G., Frenk C. S., Granato G. L., Silva L., Bressan A., Benson A. J., Cole S., 2005, MNRAS, 356, 1191
Bigiel F., Leroy A., Walter F., Brinks E., de Blok W. J. G., Madore B., Thornley M. D., 2008, AJ, 136, 2846
Blitz L., Rosolowsky E., 2006, ApJ, 650, 933
Bolatto A.D., Leroy A. K., Rosolowsky E., Walter F., Blitz L., 2008, ApJ, 686, 948
Bothwell M. S., Chapman S.C., Tacconi L.J., et al., 2010, MNRAS, in press (arXiv:0912:1598)
Bouché N., Cresci G., Davies, R., et al., 2007, ApJ, 671, 303
Bouché N., Dekel A., Genzel R., et al., 2010, ApJ, in press (arXiv:0912.1858)
Braine, J., Combes, F. 1993, A&A, 269, 7
Calzetti D. Armus, L., Bohlin R. C., Kinney A. L., Koornneef J., Storchi-Bergmann T., 2000, ApJ, 533, 682
Calzetti D., Kennicutt R. C., Engelbracht C. W et al., 2007, ApJ, 666, 870
Chabrier G., 2003, PASP, 115, 763
Chapman S. C., Blain A. W., Smail I., Ivison R. J., 2005, ApJ, 622, 772
Chapman S. C., Neri R., Bertoldi F., et al. 2008, ApJ, 689, 889
Chary R., Elbaz, D., 2001, ApJ, 556, 562
Cox, P. et al. 2006, 2007, IRAM Annual Reports, http://iram.fr/IRAMFR/ARN/AnnualReports/Years.html
Cresci G., Hicks E. K. S., Genzel R., et al., 2009, ApJ, 697, 115

Daddi E., Dickinson M., Morrison G., et al., 2007, ApJ, 670, 156
Daddi E., Dannerbauer H., Elbaz D., Dickinson M., Morrison G., Stern D., Ravindranath S., 2008, ApJ, 673, L21
Daddi E., Bournaud F., Walter F., et al., 2010, ApJ, 713, 686
Dame T. M., Hartmann D., Thaddeus P., 2001, ApJ, 547, 792
Dannerbauer H., Daddi E., Riechers D. A., Walter F., Carilli C. L., Dickinson M., Elbaz D., Morrison G. E., 2009, ApJ, 698, L178
Davé R., 2008, MNRAS, 385, 147
Davis M., Guhathakurta P., Konidaris N. P., et al. 2007, ApJ, 660, L1
de Grijs R., Parmentier G., 2007, ChJAA, 7, 155
Dekel A., Birnboim Y., 2006, MNRAS, 368, 2
Dekel A., Birnboim Y., Engel G., et al. 2009, Nature, 457, 451
Dekel A., Sari R., Ceverino D., 2009, ApJ, 703, 785
Dickman R. L., Snell R. L., Schloerb F. P., 1986, ApJ, 309, 326
Downes D., Solomon P. M., 1998, ApJ, 507, 615.
Efstathiou A., Rowan-Robinson M., 2003, MNRAS, 343, 322

Elbaz D., et al., 2010, A&A, in press
Elmegreen B., 1997, RevMexAC, 6, 165
Elmegreen B. G., 2009, in The Galaxy Disk in Cosmological Context, Proceedings of the International Astronomical Union, IAU Symposium, Volume 254. Edited by J. Andersen, J. Bland-Hawthorn, and B. Nordström, p. 289
Engel H., Tacconi L.J., Neri R., et al., 2010, ApJ, in prep.
Erb D. K., Steidel C.C., Shapley A. E., Pettini M., Reddy N. A., Adelberger K. L., 2006, ApJ, 647, 128
Erb D. K., Shapley A.E., Pettini M., Steidel C.C., Reddy N.A., Adelberger K.L., 2006b, ApJ, 644, 813
Erb D.K., 2008, ApJ, 674, 151
Förster Schreiber N. M., Genzel R., Bouché N., et al., 2009, ApJ, 706, 1364
Förster Schreiber N. M., Genzel R., Lehnert, M. D., et al., 2006, ApJ, 645, 1062
Franx M., van Dokkum P. G., Förster Schreiber N. M., Wuyts S., Labbé I., Toft S., 2008, ApJ, 688, 770
Frayer D. T., Ivison R. J., Scoville N. Z., Yun M., Evans A. S., Smail I., Blain A. W., Kneib J.-P., 1998, ApJ, 506, L7
Frayer D. T., Ivison R. J., Scoville N. Z., et al., 1999, ApJ, 514, L13
Gao Y., Solomon P. M., 2004, ApJ, 606, 271
Genzel R., Lutz D., Sturm E., et al., 1998, ApJ, 498, 579
Genzel R., Baker A. J., Tacconi L. J., Lutz D., Cox P., Guilloteau S., Omont A., 2003, ApJ, 584, 633
Genzel R., Burkert A., Bouché N., et al., 2008, ApJ, 687, 59
Goldader J. D., Joseph R. D., Doyon R., Sanders D. B., 1995, ApJ, 444, 97
Goldader J. D., Meurer G., Heckman T. M., Seibert M., Sanders D. B., Calzetti D., Steidel C. C., 2002, ApJ568, 651
Graciá-Carpio J., García-Burillo S., Planesas P., Fuente A., Usero A., 2008, A&A, 479, 703
Graciá-Carpio J., 2009, PhD Thesis
Greve T., Bertoldi F., Smail I. et al., 2005, MNRAS 259, 1165
Guilloteau S. Delannoy J., Downes D. et al., 1992, A&A, 262, 624
Iono D., Wilson C. D., Yun M. S. et al., 2009, ApJ, 695, 1537
Kaviani A., Haehnelt M. G., Kauffmann G., 2003, MNRAS, 340, 739
Kennicutt R. C., Jr., 1998a, ApJ, 498, 541
Kennicutt, R. C., Jr., 1998b, ARA&A, 36, 189
Kennicutt R. C., Jr., Calzetti D., Walter F., et al., 2007, ApJ, 671, 333
Kennicutt R. C., Jr., Hao C.-N., Calzetti D., et al., 2009, ApJ, 703, 1672
Kereš D., Katz N., Weinberg D. H., Davé R. 2005, MNRAS, 363, 2
Krumholz M.R., McKee C.F., 2005, ApJ, 630, 250
Krumholz M. R., McKee C. F., Tumlinson J., 2009, ApJ, 699, 850
Kuno N., Sato N., Nakanishi H., et al. , 2007, PASJ, 59, 117
Leroy A. K., Walter F., Brinks E., Bigiel F., de Blok W. J. G., Madore B., Thornley M. D., 2008, AJ, 136, 2782
Leroy A.K. ,Walter F., Bigiel F., et al., 2009, AJ, 137, 4670
Lutz D.,Valiante E., Sturm E., Genzel R., Tacconi L. J., Lehnert M. D., Sternberg A., Baker A.J., 2005, ApJ, 625, L83

Magnelli B., Lutz D., Berta S., et al., 2010, A&A, in press
Mauersberger R., Henkel C., Walsh W., Schulz A.,1999, A&A, 341, 256
McKee C. F., Ostriker E. C., 2007, ARA&A, 45, 565
Menéndez-Delmestre K., Blain A.W., Alexander D. M., et al. 2007, ApJ, 655, L65
Mihos, J.C. , 1999, astro-ph, 9910194
Murphy E. J., Chary R.-R., Alexander D. M., Dickinson M., Magnelli B., Morrison G., Pope A., Teplitz, H. I., 2009, ApJ, 698, 1380
Netzer H., Lutz D., Schweitzer M., et al., 2007, ApJ, 666, 806
Noeske K. G., Weiner B. J. Faber S. M., et al., 2007, ApJ, 660, L43
Nordon R., Lutz D., Shao L., et al., 2010, A&A in press
Obreschkow D., Rawlings S., 2009, MNRAS, 394, 1857
Ocvirk P., Pichon C., Teyssier R., 2008, MNRAS, 390, 1326
Omont A., Petitjean P., Guilloteau S., McMahon R. G., Solomon P. M., Pécontal, E., 1996, Nature, 382, 428
Pope A., Scott D., Dickinson M., et al., 2006, MNRAS, 370, 1185
Pope A., Chary R., Alexander D. M. et al., 2008, ApJ, 675, 1171
Reddy N. A., Erb D. K., Steidel C. C., Shapley A. E., Adelberger K. L., Pettini M., 2005, ApJ, 633, 748
Reddy N. A., Steidel C. C., 2009, ApJ, 692, 778
Reddy N. A., Erb D. K., Pettini M., Steidel C. C., Shapley, A.E., 2010, ApJ, 712, 1070
Sakamoto K., Scoville N. Z., Yun M. S., Crosas M., Genzel R., Tacconi L. J., 1999, ApJ, 514, 68
Sanders D. B., Scoville N. Z., Soifer B. T., 1991, ApJ, 370, 158
Sanders D.B., Mirabel I. F., 1996, ARA&A, 34, 749
Schmidt M., 1959, ApJ, 129, 243
Scoville N. Z., Yun M. S., Bryant P. M., 1997, ApJ, 484, 702
Shapiro K.L., Genzel R., Förster Schreiber N.M., et al., 2008, ApJ, 682, 231
Silk J., 1997, ApJ, 481, 703
Smail I. et al., 2010, in preparation
Solomon P. M., Rivolo A. R., Barrett J., Yahil, A., 1987, ApJ, 319, 730
Solomon, P.M., Sage, L.J., 1988, ApJ, 334, 613
Solomon P. M., Downes D., Radford S. J. E., Barrett J. W., 1997, ApJ, 478, 144
Steidel C. C., Shapley A. E., Pettini M., Adelberger K. L., Erb D. K., Reddy N. A., Hunt M. P., 2004, ApJ, 604, 534
Sternberg A., Hoffmann T. L., Pauldrach A. W. A., 2003, ApJ, 599, 1333
Strong A. W., Mattox J. R., 1996, A&A, 308, L21
Struck C., Kaufman M., Brinks E., Thomasson M., Elmegreen B.G., Elmegreen D.M., 2005, MNRAS, 364, 69
Tacconi L. J., Genzel R., Tecza M., Gallimore J. F., Downes D., Scoville N. Z., 1999, ApJ, 524, 732
Tacconi L. J., Neri R., Chapman S. C., et al., 2006, ApJ, 640, 228
Tacconi L. J., Genzel R., Smail I., et al., 2008, ApJ, 680, 246
Tacconi L. J., Genzel R., Neri R., et al. 2010, Nature, 463, 781
Telesco C. M., Becklin E. E., Wynn-Williams C. G., Harper D. A., 1984, ApJ, 282, 427

Toomre A., 1964, ApJ, 139, 1217

Valiante E., Lutz D., Sturm E., Genzel R., Tacconi L. J., Lehnert M. D., Baker A. J., 2007, ApJ, 660,1060
Van Dokkum P.G., 2008, ApJ, 674, 29
Veilleux S. Kim D.-C., Sanders D. B., 2002, ApJS, 143, 315
Veilleux S.,Kim D.-C., Rupke D. S. N., et al., 2009, ApJ, 701, 587
Walter F., Bertoldi F., Carilli C. et al., 2003, Nature 424, 406
Weiss A., Downes D., Walter F., Henkel C., 2007, in From z-Machines to ALMA: (Sub)Millimetre Spectroscopy of Galaxies ASP Conference Series, Vol. 375, Eds. Andrew J. Baker, Jason Glenn, Andrew I. Harris, Jeffrey G. Mangum & Min S. Yun., p.2542
Young J. S., Schloerb F. P., Kenney J. D., Lord S. D., 1986, ApJ, 304, 443
Young, J. S., Scoville N. Z., 1991, ARA&A, 29, 581

# Figures

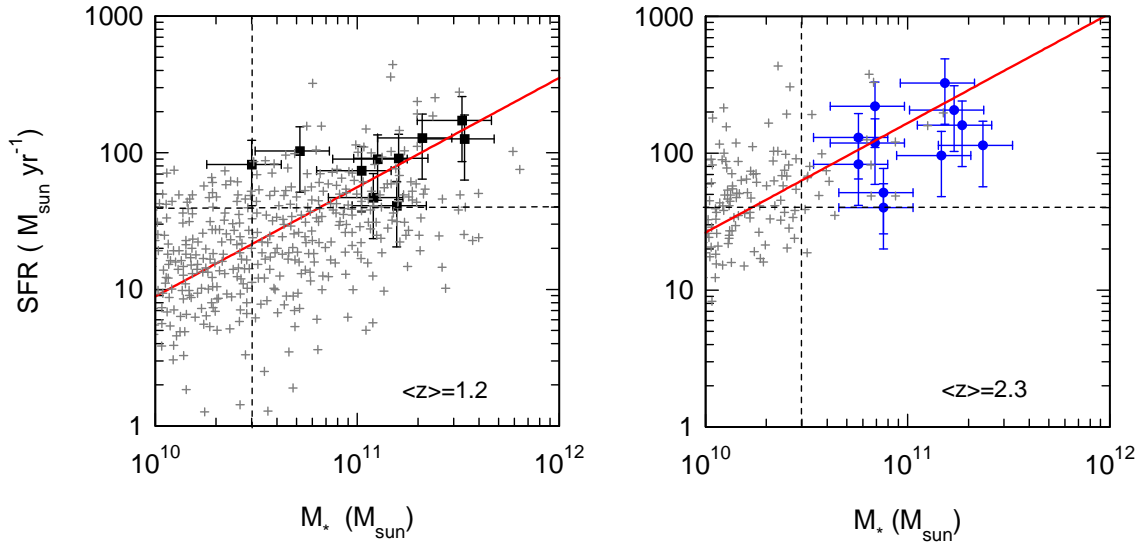

Figure 1. Properties of high-z SFGs. Shown is the location of the <z>~1.2 (left) and <z>~2.3 (right) SFGs in the stellar mass-star formation rate plane. The thick red lines denote the best fit, average $M_*$-SFR relations (the 'main-sequence line': SFR ($M_\odot$ yr$^{-1}$)=150($M_*/10^{11}$ $M_\odot$)$^{0.8}$([1+z]/3.2)$^{2.7}$, Bouché et al. 2010, Noeske et al. 2007, Daddi et al. 2007). The grey crosses are the SFR-$M_*$ data from Noeske et al. (2007) and Daddi et al. (2007) scaled to the same mean redshift as the CO observations with the $(1+z)^{2.7}$ dependence given above. The dashed vertical and horizontal lines mark the common matched selection criteria for both red-shift ranges ($M_*>4\times10^{10}$ $M_\odot$, SFR>40 $M_\odot$ yr$^{-1}$).

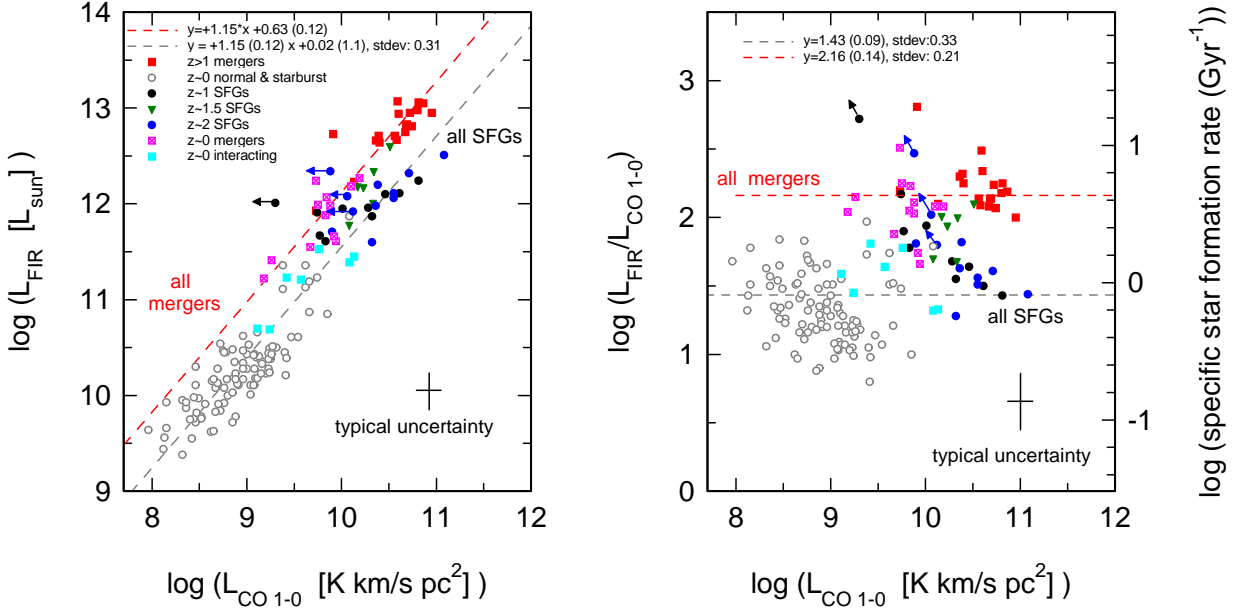

Figure 2. Correlation between observed/inferred CO 1-0 luminosity and observed/inferred FIR luminosity for different galaxy samples in our data base. Open grey circles denote isolated normal (and starburst) galaxies at z~0, from K98a, Gao & Solomon (2004), Kuno et al.(2007), Gracia-Carpio et al. (2008), Gracia-Carpio (2009) and Leroy et al. (2008, 2009); filled cyan squares denote interacting z~0 galaxies from the same references; filled blue circles are <z>=2.3 SFGs (BX) and filled black circles <z>=1.2 SFGs (EGS), both from Tacconi et al. (2010, in prep.); and filled green triangles are <z>=1.5 SFGs (BzK) from Daddi et al. (2010). Crossed magenta squares are z~0 LIRG/ULIRG mergers from K98a, Gracia-Carpio et al. (2008) and Gracia-Carpio (2009). Red squares are z=1-3.5 SMGs from Greve et al. (2005), Engel et al (2010) and Smail et al. (in prep). In cases where rotationally excited CO lines were observed correction factors discussed in section 2.6 have been applied to the observed luminosities (see also Table 1). The typical total (statistical plus systematic) 1σ uncertainty is shown as a large black cross in the lower right of the panels. Left panel: luminosity-luminosity correlation. Dotted grey and red lines give the results of the fits to the SFGs and luminous mergers, respectively, including in each case all red-shifts. The fits assign equal weight to all data points. Right panel: $L_{FIR}/L_{CO}$ as a function of $L_{CO}$. Dotted grey and red lines give the results of average values to the SFGs and luminous mergers, respectively, including in each case all red-shifts. The fits assign equal weight to all data points, and uncertainties

in brackets are 3σ formal fit errors. The specific star formation rate (SFR/$M_{mol-gas}$) computed from $L_{FIR}/L_{CO}$ is given on the right vertical axis.

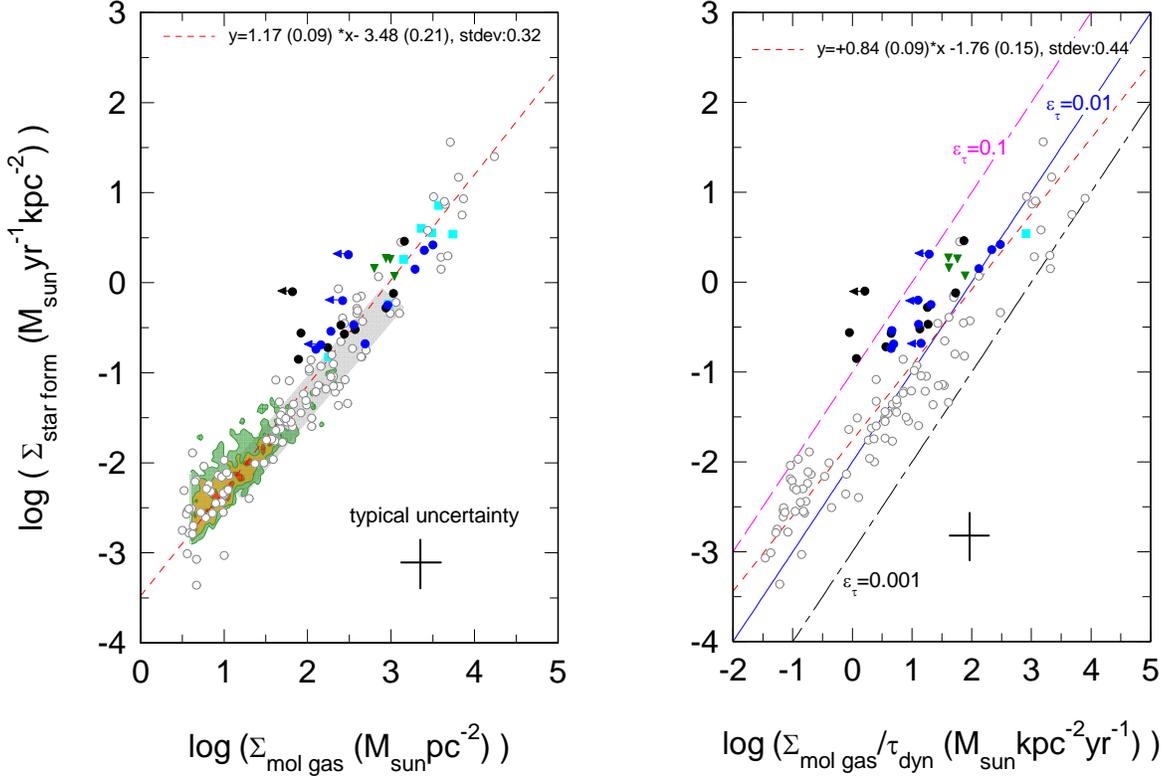

Figure 3. Molecular surface density relations for SFGs. Symbols and data base are the same as in Figure 2 but there are fewer galaxies than in Figure 2 because of the lack of availability of size measurements. For the high-z SFGs surface densities are calculated by dividing 50% of the integrated star formation rate and molecular gas mass (including a 36% correction for helium) by the effective area of the half-light radius ($\pi R_{1/2}^2$). Left panel: $\Sigma_{molgas}$-$\Sigma_{starform}$ relation ('Kennicutt-Schmidt'-relation). In addition to our data base, we have also included in green contours with green/orange/red shading the average distribution of the spatially resolved data in normal galaxies given in Bigiel et al. (2008, green/orange colours). The grey shaded region denotes the relation found from spatially resolved observations in M51 (Kennicutt et al. 2007). Right panel: $\Sigma_{molgas}/\tau_{dyn}$-$\Sigma_{starform}$ relation ('Elmegreen-Silk' relation). The fits assign equal weight to all data points, and uncertainties in brackets are 3σ formal fit errors. The crosses in the bottom right denote the typical total (statistical + systematic) 1σ uncertainties. The black dash-dotted, blue continuous and magenta long-dashed lines mark lines of (constant) 0.1%, 1% and 10% efficiency per dynamical time scale $\varepsilon_\tau$ (equation 4).

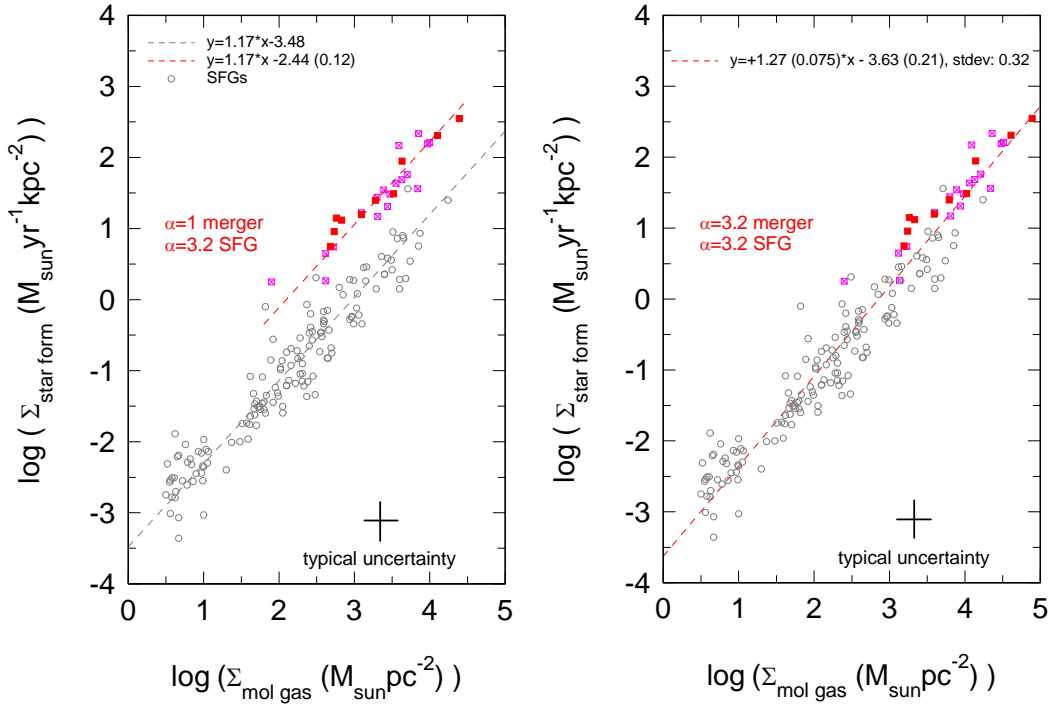

Figure 4. Molecular Kennicutt-Schmidt surface density relation for luminous z~0 and z~1-3.5 mergers (z~0 LIRGs/ULIRGs: magenta squares, z≥1 SMGs: red squares). The left panel shows their location in the KS-plane along with the SFGs (at all z, open grey circles) from Figure 3 if the a priori best conversion factors for SFGs ($\alpha=\alpha_G$) and mergers ($\alpha=\alpha_G/3.2$) are chosen. The right panel shows the same plot for the choice of a universal conversion factor of $\alpha=\alpha_G$ for all galaxies in the data base. This was the choice in the K98a paper but leads to a significant overestimate of gas fractions in almost all major mergers. The fits assign equal weight to all data points and uncertainties in brackets are $3\sigma$ formal fit errors. The crosses in the lower right denote the typical total (statistical + systematic) $1\sigma$ uncertainty.

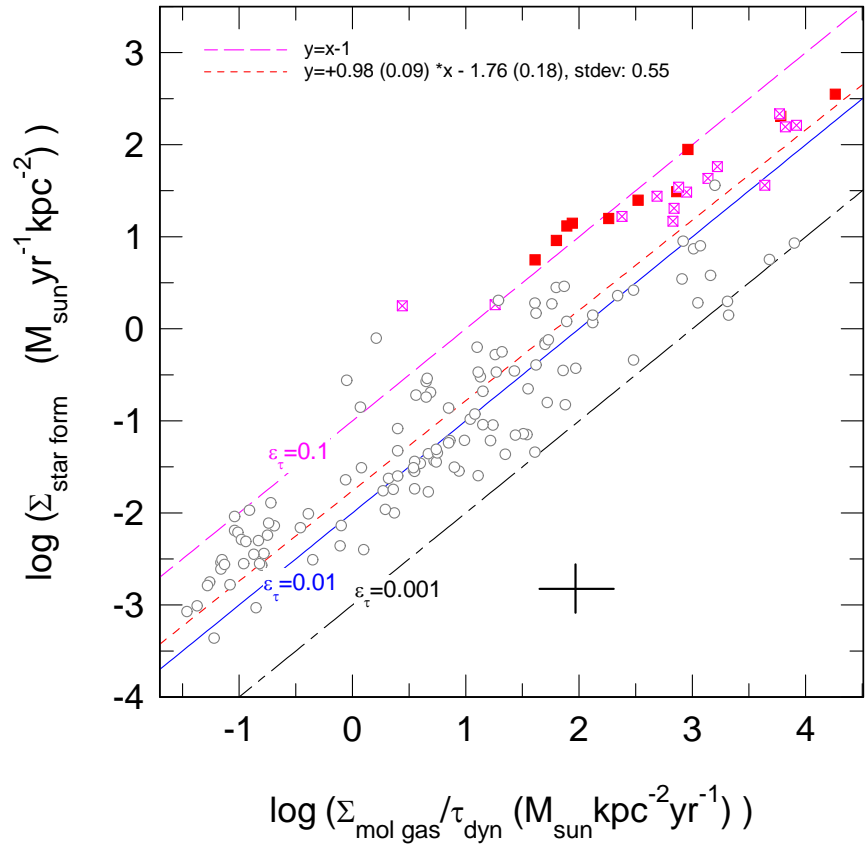

Figure 5. Molecular 'Elmegreen-Silk'-relation for mergers and SMGs. Symbols are the same as in Figures 2-4. Here the a priori best conversion factors for SFGs ($\alpha=\alpha_G$) and mergers ($\alpha=\alpha_G/3.2$) are chosen. The short-dashed red line is the best fit to all data. The fits assign equal weight to all data points and the uncertainties (in brackets) are $3\sigma$ formal fit errors. The cross at the bottom denotes the typical total (statistical + systematic) $1\sigma$ uncertainty. As in Figure 3, the black dash-dotted, blue continuous and magenta long-dashed lines mark lines of (constant) 0.1%, 1% and 10% efficiency per dynamical time scale $\varepsilon_\tau$ (equation 4).

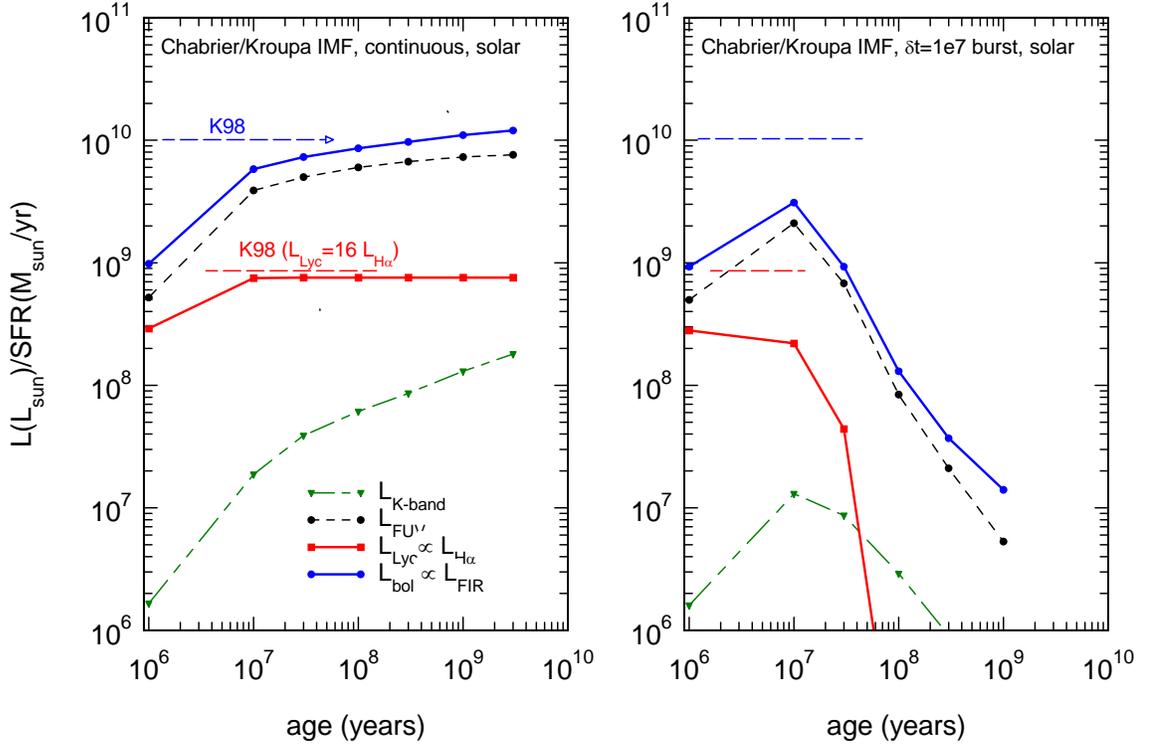

Figure 6. Dependence of different star formation tracers on star formation histories. Left panel: Ratio of luminosity in a given tracer (Hα, far-infrared (=total bolometric), far-UV and near-infrared luminosities, see legend) to the star formation rate for a constant star formation rate, as a function of time. Right panel: the same quantities for a moderately short duration ($\delta t=10^7$ years) burst model (SFR(t)=SFR$_0$exp(-t/δt)). In both panels we have used the STARS stellar population synthesis code (Sternberg et al. 2003), with solar metallicity templates and a Kroupa (~Chabrier) IMF. In both panels the dashed blue and red horizontal lines mark the values of the K98a calibrations of the far-IR and Hα indicators (corrected to a Chabrier/Kroupa IMF). These plots show that these calibrations are applicable for the equilibrium (t>>$10^7$ yrs) in constant star formation histories but not for smaller ages for the far-IR indicator and not for any age for short duration bursts post the peak of the burst. In all these cases the indicators underestimate the intrinsic peak star formation rate (SFR$_0$).

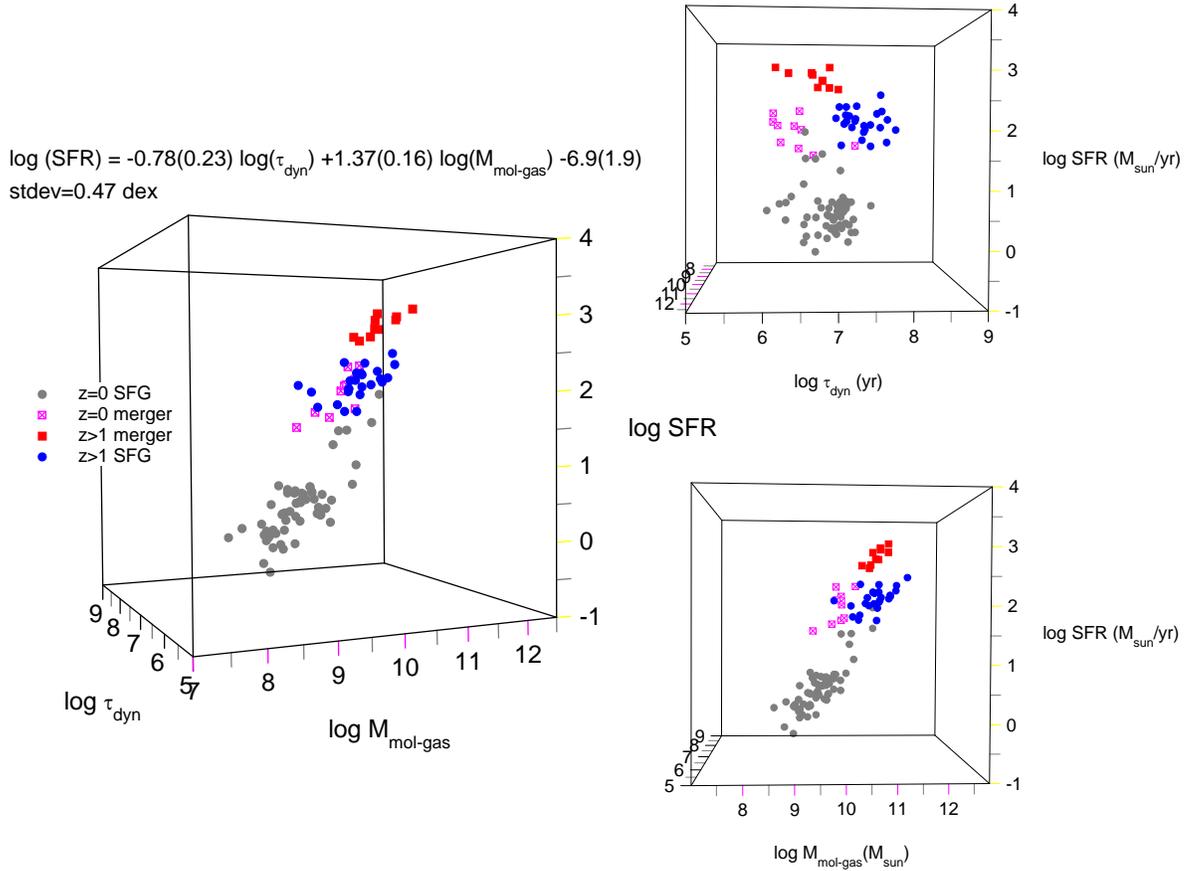

Figure 7. Relation between star formation rate (vertical axis), molecular gas mass and dynamical time in a three-dimensional plot. Filled grey and blue circles are z~0 and z≥1 SFGs, crossed open magenta squares are z~0 LIRG/ULIRG mergers and filled red squares are z≥1 SMGs. The bottom right panel (the $M_{mol-gas}$-SFR projection) is essentially the same as Figure 2 and shows the clear separation of mergers and SFGs (in all three panels we use $α_{merger}$=1 and $α_{SFG}$=3.2). The top right panel shows the $τ_{dyn}$-SFR projection, which again shows a clear separation between compact mergers on the left and extended SFGs on the right. For the high-z SFGs and mergers (which are of comparable dynamical, stellar mass and gas fraction: Tacconi et al. 2008, 2010) there appears to be a clear sequence of star formation rate on dynamical time (slope -0.8). The central left panel shows that there exists a three-dimensional projection of the data in which differences between different types of galaxies are minimal. The best fitting scaling relation (log(SFR)=-0.78 (±0.23) log($τ_{dyn}$) +1.37 (±0.16) log($M_{mol-gas}$) -6.9(±1.9)) has a scatter of ±0.47 dex.

**Table 1. Observed and Derived Properties of Spatially Resolved z≥1 SFGs and SMGs**

| source[1] | z | $v_d^2$ | $R_{1/2}{}^3$ | SFR[4] | $F_J(CO)^5$ | $\sigma(F_J)^5$ | $L_J(CO)^6$ | $\sigma(L_J)$ | $M_{mol-gas}{}^7$ | $M_*{}^8$ | $\log(\Sigma_{mol-gas})^9$ | $\log(\Sigma_{mol-gas}/\tau_{dyn})^9$ | $\log(\Sigma_{star-form})^9$ |
|---|---|---|---|---|---|---|---|---|---|---|---|---|---|
| | | km/s | kpc | $M_\odot$/yr | Jy km/s | | K km/s pc$^2$ | | $M_\odot$ | $M_\odot$ | $M_\odot$pc$^{-2}$ | $M_\odot$yr$^{-1}$kpc$^{-2}$ | $M_\odot$yr$^{-1}$kpc$^{-2}$ |
| EGS13004291 | 1.20 | 150 | 7.2 | 172 | 3.70 | 0.15 | 3.2E+10 | 1.3E+09 | 2.8E+11 | 3.3E+11 | 2.94 | 1.26 | -0.28 |
| EGS12007881 | 1.17 | 174 | 8.7 | 91 | 1.15 | 0.06 | 9.4E+09 | 4.9E+08 | 8.3E+10 | 1.6E+11 | 2.24 | 0.56 | -0.72 |
| EGS13017614 | 1.18 | 223 | 6.3 | 74 | 1.25 | 0.10 | 1.1E+10 | 8.4E+08 | 9.3E+10 | 1.1E+11 | 2.57 | 1.13 | -0.52 |
| EGS13035123 | 1.12 | 139 | 8.6 | 126 | 1.90 | 0.05 | 1.4E+10 | 3.8E+08 | 1.3E+11 | 3.4E+11 | 2.44 | 0.65 | -0.57 |
| EGS13004661 | 1.19 | 73 | 6.9 | 82 | 0.32 | 0.06 | 2.8E+09 | 5.2E+08 | 2.4E+10 | 3.0E+10 | 1.92 | -0.05 | -0.56 |
| EGS13003805 | 1.23 | 255 | 5.2 | 128 | 2.25 | 0.15 | 2.0E+10 | 1.4E+09 | 1.8E+11 | 2.1E+11 | 3.03 | 1.73 | -0.12 |
| EGS12011767 | 1.28 | 107 | 7.3 | 47 | 0.30 | 0.06 | 2.9E+09 | 5.9E+08 | 2.6E+10 | 1.2E+11 | 1.89 | 0.07 | -0.85 |
| EGS12012083 | 1.12 | 110 | 4.6 | 103 | -0.13 | 0.04 | -9.9E+08 | 3.3E+08 | -8.7E+09 | 5.2E+10 | 1.82 | 0.21 | -0.10 |
| EGS13011439 | 1.10 | 114 | 2.2 | 90 | 0.70 | 0.15 | 5.1E+09 | 1.1E+09 | 4.5E+10 | 1.3E+11 | 3.16 | 1.87 | 0.46 |
| EGS13011148 | 1.17 | 316 | 4.3 | 41 | 0.41 | 0.08 | 3.4E+09 | 6.6E+08 | 3.0E+10 | 1.6E+11 | 2.40 | 1.27 | -0.46 |
| HDF-BX1439 | 2.19 | 265 | 8.0 | 83 | -0.23 | 0.08 | -6.6E+09 | 2.2E+09 | -5.9E+10 | 5.7E+10 | 2.16 | 0.69 | -0.69 |
| Q1623-BX599 | 2.33 | 265 | 2.8 | 130 | 0.60 | 0.10 | 1.8E+10 | 3.0E+09 | 1.6E+11 | 5.7E+10 | 3.50 | 2.48 | 0.42 |
| Q1623-BX663 | 2.43 | 256 | 5.5 | 119 | -0.18 | 0.06 | -5.7E+09 | 1.9E+09 | -5.1E+10 | 6.9E+10 | 2.42 | 1.10 | -0.20 |
| Q1700-MD69 | 2.29 | 217 | 9.4 | 160 | 0.42 | 0.06 | 1.2E+10 | 1.7E+09 | 1.1E+11 | 1.9E+11 | 2.28 | 0.66 | -0.54 |
| Q1700-MD94 | 2.34 | 217 | 9.6 | 326 | 2.00 | 0.30 | 6.0E+10 | 9.0E+09 | 5.3E+11 | 1.5E+11 | 2.96 | 1.32 | -0.25 |
| Q1700-MD174 | 2.34 | 240 | 3.6 | 114 | 0.60 | 0.08 | 1.8E+10 | 2.4E+09 | 1.6E+11 | 2.4E+11 | 3.29 | 2.12 | 0.15 |
| Q1700-BX691 | 2.19 | 238 | 6.7 | 52 | 0.15 | 0.05 | 4.0E+09 | 1.2E+09 | 3.5E+10 | 7.6E+10 | 2.10 | 0.65 | -0.74 |
| Q2343-BX389 | 2.17 | 259 | 4.2 | 220 | -0.15 | 0.05 | -3.8E+09 | 1.3E+09 | -3.3E+10 | 6.9E+10 | 2.49 | 1.29 | 0.31 |
| Q2343-BX442 | 2.18 | 238 | 6.7 | 96 | 0.43 | 0.08 | 1.1E+10 | 2.1E+09 | 1.0E+11 | 1.5E+11 | 2.55 | 1.11 | -0.47 |
| Q2343-BX610 | 2.21 | 324 | 3.8 | 207 | 0.95 | 0.08 | 2.6E+10 | 2.2E+09 | 2.3E+11 | 1.7E+11 | 3.40 | 2.34 | 0.36 |
| Q2343-MD59 | 2.01 | 157 | 5.5 | 40 | 0.46 | 0.10 | 1.0E+10 | 2.3E+09 | 9.2E+10 | 7.6E+10 | 2.69 | 1.15 | -0.68 |
| SMMJ02399-0136 | 2.81 | 590 | 5.1 | 1147 | 1.26 | 0.16 | 4.9E+10 | 6.3E+09 | 8.8E+10 | | 2.73 | 1.80 | 0.84 |
| SMMJ09431+4700 | 3.35 | 295 | 1.4 | 873 | 0.85 | 0.13 | 2.5E+10 | 3.8E+09 | 5.5E+10 | | 3.63 | 2.96 | 1.83 |
| SMMJ105141+5719 | 1.21 | 457 | 3.1 | 648 | 2.61 | 0.55 | 5.0E+10 | 1.1E+10 | 3.4E+10 | | 2.76 | 1.94 | 1.04 |
| SMMJ123549+6215 | 2.20 | 442 | 0.9 | 897 | 1.55 | 0.16 | 4.0E+10 | 4.1E+09 | 7.1E+10 | 1.2E+11 | 4.10 | 3.78 | 2.20 |
| SMMJ123634+6212 | 1.22 | 343 | 4.1 | 465 | 1.75 | 0.30 | 3.4E+10 | 5.9E+09 | 5.2E+10 | | 2.68 | 1.61 | 0.63 |
| SMMJ123707+6214 | 2.49 | 317 | 2.8 | 508 | 0.59 | 0.12 | 1.9E+10 | 3.8E+09 | 3.4E+10 | 1.2E+11 | 2.83 | 1.89 | 1.01 |
| SMMJ131201+4242 | 3.41 | 430 | 3.0 | 670 | 1.00 | 0.20 | 3.0E+10 | 6.1E+09 | 6.7E+10 | | 3.09 | 2.26 | 1.09 |
| SMMJ131232+4239 | 2.33 | 346 | 2.0 | 508 | 0.99 | 0.22 | 2.8E+10 | 6.3E+09 | 5.0E+10 | | 3.28 | 2.52 | 1.29 |

| | | | | | | | | | | | | |
|---|---|---|---|---|---|---|---|---|---|---|---|---|
| SMMJ163650+4057 | 2.39 | 523 | 2.4 | 886 | 2.30 | 0.20 | 6.9E+10 | 6.0E+09 | 1.2E+11 | 2.3E+11 | 3.52 | 2.86 | 1.37 |
| SMMJ163658+4105 | 2.45 | 590 | 0.8 | 1124 | 1.80 | 0.15 | 5.6E+10 | 4.7E+09 | 1.0E+11 | 2.6E+11 | 4.39 | 4.26 | 2.44 |
| BzK4171 | 1.47 | 256 | 3.7 | 103 | 0.65 | 0.08 | 1.8E+10 | 2.3E+09 | 9.4E+10 | 6.5E+10 | 3.04 | 1.89 | 0.08 |
| BzK210000 | 1.52 | 199 | 4.3 | 220 | 0.64 | 0.07 | 2.0E+10 | 2.1E+09 | 1.0E+11 | 6.5E+10 | 2.94 | 1.61 | 0.28 |
| BzK16000 | 1.52 | 260 | 4.1 | 152 | 0.42 | 0.06 | 1.3E+10 | 1.8E+09 | 6.6E+10 | 6.5E+10 | 2.80 | 1.62 | 0.17 |
| BzK17999 | 1.41 | 207 | 3.6 | 148 | 0.57 | 0.07 | 1.5E+10 | 1.9E+09 | 7.7E+10 | 6.5E+10 | 2.99 | 1.76 | 0.27 |

[1] EGS <z>=1.2 are SFGs drawn from the AEGIS survey (Davis et al. 2007, Noeske et al. 2007, Tacconi et al. 2010, in prep.); BX/MD<z>=2.3 SFGs are drawn from the Erb et al. (2006) Hα sample of BX-galaxies (Steidel et al. 2004, Adelberger et al. 2005); SMM are z=1-3.5 submillimeter galaxies from Greve et al. (2005), Tacconi et al. (2006, 2008), Engel et al. (2010) and Smail et al. (in prep.); BzK are <z>=1.5 SFGs from Daddi et al. (2010).

[2] maximum circular velocity, 1σ uncertainty typically 30-40%

[3] HWHM or half-light radius obtained from fits to Hα (for BX: Erb et al. 2006, Förster Schreiber et al. 2009), optical/UV stellar light (for EGS: Cooper priv. comm., for BzK: Daddi et al. 2010) and (where available) from CO maps (for z≥1 SMG: Tacconi et al. 2006, 2008, Engel et al. 2010, for EGS/BX: Tacconi et al. 2010, in prep., for BzK : Daddi et al. 2010). When several indicators were available the number quoted is an average.

[4] extinction corrected star formation rate for a Chabrier (2003) IMF. Star formation rates were estimated in the following way: z≥1 SMGs: from 850μm flux and the Pope et al. (2006) and Magnelli et al. (2010) calibrations ($L_{FIR}$=1.2 (+0.8,-0.3) x$10^{12}$ $S_{850\mu m}$(Jy) ($L_\odot$)) and with the K98a conversion from luminosity to star formation rate (SFR=1.3x$10^{-10}$ $L_{FIR}$, corrected to Chabrier IMF, where the factor 1.3 is a correction from FIR to total IR (8-1000μm luminosity (Gracia-Carpio et al. 2008), Figure 5);
EGS: from a combination of extinction corrected Hα/[OII]/GALEX UV and 24μm Spitzer luminosities (extrapolating to the FIR with Chary & Elbaz 2001 SEDs) (Noeske et al. 2007); BX: from extinction corrected Hα-luminosities (Erb et al. 2006, Förster Schreiber et al. 2009), using the E(B-V) reddening obtained from the UV-SEDs (Erb et al. 2006) and the Calzetti et al. (2000) recipe (A(Hα)=7.5 E(B-V) ), with SFR=(L(Hα)$_0$/2.1x$10^{41}$erg/s) (K98b);
BzK: from a combination of extinction corrected UV-luminosities and 24μm Spitzer luminosities, extrapolating to the FIR with Chary & Elbaz 2001 SEDs) (Daddi et al. 2007, 2010); typical systematic uncertainties are ±50%.

[5] Observed CO J- (J-1) line integrated flux (Jy km/s) and 1σ error (corrected for lensing magnification where necessary). For EGS and BX galaxies J=3, for z≥1 SMG J=4,3,2 depending on red-shift, for BzK J=2. Negative values in this column and following columns denote 3σ upper limits.

[6] CO J- (J-1) integrated line luminosity (Solomon et al. 1997): $L_J$ (CO, K km/s pc$^2$)=$3.25 \times 10^{13} F_J(Jy)[\nu_{obs}(GHz)]^{-2} [D_L(Gpc)]^2 [1+z]^{-3}$, where $\nu_{obs}$ is the observed line frequency in GHz and $D_L$ is the luminosity distance in Gpc.

[7] molecular gas mass: $M_{mol-gas}=1.36 \alpha R_{1J} L_J(CO)$ (M$_\odot$), where α is the conversion factor (3.2 for SFGs, 1 for z≥1 SMGs) and $R_{1J}=L'_{CO\ 1-0}/L'_{CO\ J-(J-1)}$=1.2 and 2 for J=2 and 3 SFGs. For the z≥1 SMGs we use $R_{1J}$=1.1, 1.3 and 1.6 for J=2,3 and 4. Conversion factors are probably uncertain by ±30% for SFGs and a factor of at least ±50% for z~0 ULIRG/ z≥1 SMGs.

[8] stellar masses for a Chabrier IMF, obtained from SED population synthesis fitting to the combined UV/optical/near-IR SEDs (Erb et al. 2006, Förster Schreiber et al. 2009, Noeske et al. 2007, Daddi et al. 2007, 2010); uncertainty ±40%.

[9] surface density within $R_{1/2}$, determined, for instance, as: $\Sigma_{mol-gas}=0.5\ M_{mol-gas}/(\pi R_{1/2}^2)$